\documentclass[conference]{IEEEtran}
\IEEEoverridecommandlockouts
\usepackage{cite}
\usepackage{amsmath,amssymb,amsfonts}
\usepackage{algorithm}
\usepackage{algorithmic}
\usepackage{graphicx}
\usepackage{textcomp}
\usepackage{xcolor}
\usepackage{multirow}
\usepackage{booktabs}
\usepackage{adjustbox}
\usepackage{pifont}
\usepackage{color} 
\def\BibTeX{{\rm B\kern-.05em{\sc i\kern-.025em b}\kern-.08em
    T\kern-.1667em\lower.7ex\hbox{E}\kern-.125emX}}

\makeatletter
\newcommand{\linebreakand}{%
    \end{@IEEEauthorhalign}
    \hfill\mbox{}\par
    \mbox{}\hfill\begin{@IEEEauthorhalign}
}
\makeatother
    
\begin{document}
\title{Meta-optimized Joint Generative and Contrastive Learning for Sequential Recommendation}

% \DeclareRobustCommand*{\IEEEauthorrefmark}[1]{%
%   \raisebox{0pt}[0pt][0pt]{\textsuperscript{\footnotesize #1}}%
% }
% \author{
%     \IEEEauthorblockN{
%         \IEEEauthorrefmark{1},
%         \IEEEauthorrefmark{1}, 
%         \IEEEauthorrefmark{1},
%         \IEEEauthorrefmark{1},
%         \IEEEauthorrefmark{2}, \\
%         \IEEEauthorrefmark{3}
%        and \IEEEauthorrefmark{4} 
%     }
%     \IEEEauthorblockA{
%         \IEEEauthorrefmark{1}School of Computer Science and Technology, Soochow University, Suzhou, China\\
%         \IEEEauthorrefmark{2}Department of Computing, Macquarie University, Australia \\
%         \IEEEauthorrefmark{3}Texas Tech University, USA \\
%         \IEEEauthorrefmark{4}The Hong Kong University of Science and Technology, China\\
%         yjhaozb@stu.suda.edu.cn\  
%         \{ppzhao, jhfang, jfqu\}@suda.edu.cn \\
%         guanfeng.liu@mq.edu.au \    
%         victor.sheng@ttu.edu \    
%         zxf@cse.ust.hk
%     }
% }

\author{\IEEEauthorblockN{Yongjing Hao}
\IEEEauthorblockA{\textit{Soochow University} \\
Suzhou, China \\
yjhaozb@stu.suda.edu.cn}
\and
\IEEEauthorblockN{Pengpeng Zhao}
\IEEEauthorblockA{\textit{Soochow University} \\
Suzhou, China \\
ppzhao@suda.edu.cn}
\and
\IEEEauthorblockN{Junhua Fang}
\IEEEauthorblockA{\textit{Soochow University} \\
Suzhou, China \\
jhfang@suda.edu.cn}
\and
\IEEEauthorblockN{Jianfeng Qu}
\IEEEauthorblockA{\textit{Soochow University} \\
Suzhou, China \\
jfqu@suda.edu.cn} 
\linebreakand
\IEEEauthorblockN{Guanfeng Liu}
\IEEEauthorblockA{\textit{Macquarie University} \\
Sydney, Australia \\
guanfeng.liu@mq.edu.au}
\and
\IEEEauthorblockN{Fuzhen Zhuang}
\IEEEauthorblockA{\textit{Beihang University} \\
Beijing, China \\
zhuangfuzhen@buaa.edu.cn}
\and
\IEEEauthorblockN{Victor S. Sheng}
\IEEEauthorblockA{\textit{Texas Tech University} \\
Lubbock, USA \\
victor.sheng@ttu.edu}
\and
\IEEEauthorblockN{Xiaofang Zhou}
\IEEEauthorblockA{\textit{The Hong Kong University} \\
\textit{of Science and Technology} \\
HongKong SAR, China \\
zxf@cse.ust.hk}
}
\maketitle

\begin{abstract}
Sequential Recommendation (SR) has received increasing attention due to its ability to capture user dynamic preferences. Recently, Contrastive Learning (CL) provides an effective approach for sequential recommendation by learning invariance from different views of an input. However, most existing data or model augmentation methods may destroy semantic sequential interaction characteristics and often rely on the hand-crafted property of their contrastive view-generation strategies. In this paper, we propose a Meta-optimized Seq2Seq Generator and Contrastive Learning (Meta-SGCL) for sequential recommendation, which applies the meta-optimized two-step training strategy to adaptive generate contrastive views. Specifically, Meta-SGCL first introduces a simple yet effective augmentation method called Sequence-to-Sequence (Seq2Seq) generator, which treats the Variational AutoEncoders (VAE) as the view generator and can constitute contrastive views while preserving the original sequence's semantics. Next, the model employs a meta-optimized two-step training strategy, which aims to adaptively generate contrastive views without relying on manually designed view-generation techniques. Finally, we evaluate our proposed method Meta-SGCL using three public real-world datasets. Compared with the state-of-the-art methods, our experimental results demonstrate the effectiveness of our model and the code is available\footnote{https://anonymous.4open.science/status/Meta-SGCL-05B5}.
\end{abstract}

\begin{IEEEkeywords}
Sequential Recommendation, Seq2Seq Generator, Contrastive Learning, Meta-optimized
\end{IEEEkeywords}

\section{Introduction}
Recommender System (RS) have been proven to be useful tools to improve users’ experience by providing personal advice and helping users to deal with the so-called information overload \cite{DBLP:conf/sigir/Wang0WFC19,DBLP:journals/corr/abs-2305-04322,DBLP:conf/icde/ZhaoZ0LS021}. In real-world scenarios, users’ preferences are intrinsically dynamic and change over time while the next item depends largely on the items the user has engaged in recently. Hence, Sequential Recommendation Models (SRM) \cite{DBLP:conf/icde/ChenYNP0020} have been proposed to explicitly capture the sequential dependencies within user-item interaction sequences for providing more accurate recommendation \cite{DBLP:conf/ijcai/WangHWCSO19,DBLP:conf/wsdm/ChenXZT0QZ18,DBLP:journals/tnn/ZhangCSN22}, and have been successfully applied in various online applications \cite{DBLP:conf/aaai/YuZLZ19}.

Increasing research interests have been put in sequential recommendation systems with a number of models which have been proposed. Earlier SRMs utilize Markov Chain (MC) \cite{he2016fusing} and Recurrent Neural Network (RNN) \cite{zhao2020go} to mine users’ dynamic preferences. MC-based SRMs such as Factorizing Personalized Markov Chain (FPMC) \cite{DBLP:conf/www/RendleFS10} factorize the user-item transition matrix to learn the short-term transition patterns and perform well in high-sparsity scenarios \cite{de2021transformers4rec,DBLP:conf/recsys/WuLHS20}. Another line of work adopts RNN and its variants (e.g., Gated Recurrent Unit (GRU) \cite{srnn2016} and Long Short-Term Memory (LSTM) \cite{zhu2017next}) to extract long-term user preference and works well in dense settings \cite{DBLP:conf/cikm/HidasiK18}. Nowadays, Transformer \cite{DBLP:conf/nips/VaswaniSPUJGKP17} has shown promising results in many fields such as Computer Vision (CV) \cite{DBLP:conf/iccv/LiuL00W0LG21,DBLP:conf/iclr/DosovitskiyB0WZ21} and Natural Language Processing (NLP) \cite{DBLP:conf/naacl/DevlinCLT19}. Self-Attention Network (SAN), the key component of the transformer, has shown its strong ability in sequence modeling. In the light of self-attention mechanism, many SAN-based SRMs are also proposed to tackle the sequential recommendation problem and significantly improve the models’ performance \cite{DBLP:conf/recsys/WuLHS20,DBLP:conf/wsdm/LiWM20,DBLP:conf/cikm/HeZ0WKM21}. The ability to model short- and long-term dependencies over user-item interaction sequence simultaneously through the self-attention mechanism is the key to the success of SAN-based SRMs \cite{kang18attentive}. 

Recently, Self-Supervised Learning (SSL) has achieved promising results in generating representations using small labeled data by yielding auxiliary self-supervision signals in CV \cite{DBLP:conf/icml/ChenK0H20,DBLP:conf/iclr/HjelmFLGBTB19} and NLP \cite{DBLP:conf/naacl/DevlinCLT19}. Motivated by this, some works introduce Contrastive Learning (CL) into sequential recommendations to cope with data sparsity issues and dig out supervised signals from the data itself.
S$^3$-Rec \cite{DBLP:conf/cikm/ZhouWZZWZWW20} utilizes the correlations among attributes, items, and sub-sequences through mutual information maximization. CLS4Rec \cite{DBLP:journals/corr/abs-2010-14395} uses three random data augmentation operations (i.e., item crop, item mask, and item reorder) to generate contrastive views for sequential recommendation. Next, CoSeRec \cite{DBLP:journals/corr/abs-2108-06479} proposes two data augmentation operators (i.e., informative substitute and insert) to extend CLS4Rec.  
To generate informative augmentations for user behavior sequences, CCL \cite{DBLP:conf/cikm/BianZZCHYW21} employs a learnable context-aware generator for data augmentation, and performs CL on different augmented samples.
More recently, DuoRec \cite{DBLP:conf/wsdm/QiuHYW22} applies model augmentation random drop neurons operator to generate views for contrastive learning.  Then, besides the random drop neurons, SRMA \cite{yu2022self} also proposes random layer drop and encoder complement model augmentation operations for sequential contrastive learning.

% \begin{figure}[t]
% \centering
% \includegraphics[width=0.64\textwidth]{fig/FIG2.pdf}
% \vspace{-1.35cm}
% \caption{An example of data-based augmentation operations, including random item crop, item mask, item reorder, informative substitute, and insert.} 
% % \vspace{-0.5cm}
% \label{fig1}
% \end{figure}

% \begin{figure}[t]
% \centering
% \includegraphics[width=0.57\textwidth]{fig/FIG3.pdf}
% \vspace{-2.8cm}
% \caption{Two model-based augmentation operations, including random drop neurons and layers.} 
% % \vspace{-0.5cm}
% \label{fig2}
% \end{figure}
\begin{figure}[t]
\centering
\includegraphics[width=0.53\textwidth]{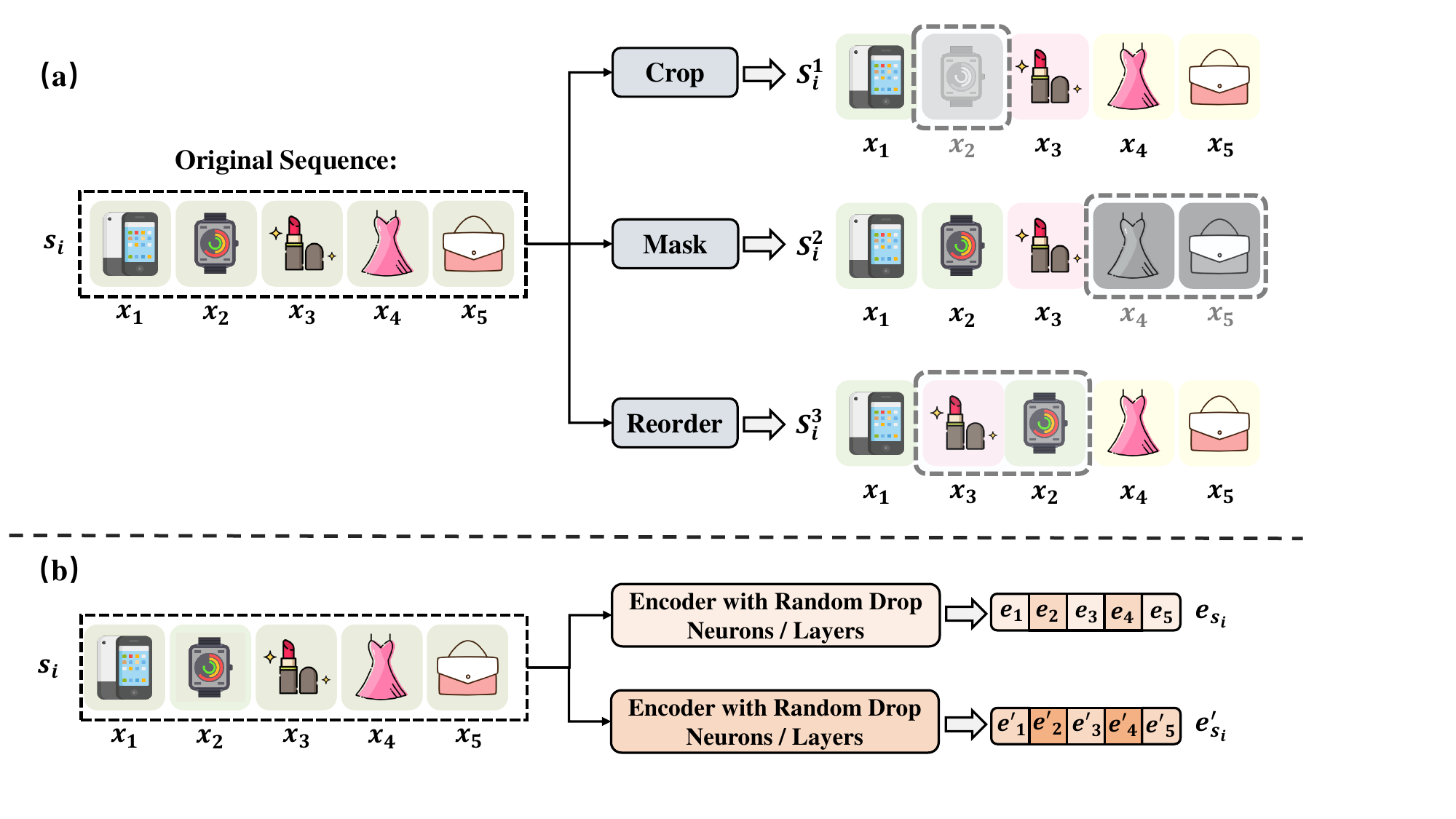}
\vspace{-0.8cm}
\caption{Motivating examples of augmentations in existing CL-based SR: (a) data-based augmentation and (b) model-based augmentation.} 
% \caption{(a) Data-based augmentation, (b) model-based augmentation.} 
% \vspace{-0.5cm}
\label{fig1}
\end{figure}

Although many augmentation strategies have been explored, the existing methods often use a manually defined way to generate contrast views, and in addition, the above augmentation strategies might disrupt original semantics and sequential properties for recommendation \cite{DBLP:journals/corr/abs-2010-14395}. 
As shown in Figure \ref{fig1}(a), we assume that a user behavior sequence is: $s_i=[x_1, x_2, x_3, x_4, x_5]$, and the augmentation views after random item crop, random item mask and random item reorder are $s^1_i=[x_1, x_3,x_4, x_5]$, $s^2_i=[x_1, x_2, x_3, [\mathrm{mask}],[\mathrm{mask}]]$ and $s^3_i=[x_1, x_3,x_2,x_4, x_5]$. 
We argue that some essential sequential correlations of $s_i$ may be disturbed in augmentation views $s^1_i, s^2_i$ and $s^3_i$. 
On the other hand, some model-based augmentation methods such as random drop neurons and random drop layers\cite{yu2022self} might also lose the original semantics information. As shown in Figure \ref{fig1}(b), the $\boldsymbol{e}_{s_i}$ is the embedding of sequence $s_{i}$, and the augmentation view after random drop neurons or layers can obtain new embedding $\boldsymbol{e}'_{s_i}$ for the same sequence $s_{i}$. 
We argue that existing model-based augmentation methods discard essential neurons or layers, and are insufficient to generate optimal contrastive views, leading to a suboptimal recommendation performance. 

A natural way to tackle this problem is to adaptively generate the augmentation views for contrastive learning without randomly disturbing the information of data or models. Existing generative recommendation models can be broadly categorized into GAN-based methods \cite{DBLP:journals/tnn/GuoZCLXCDH22} and VAE-based methods \cite{VAE}. GAN-based methods utilize adversarial training \cite{DBLP:conf/www/WangYCZWZL22} to optimize the generator for predicting user interactions, while VAE-based methods \cite{DBLP:conf/nips/MaZ0Y019} involve learning an encoder for posterior estimation and a decoder for predicting interaction probabilities for all items. Compared with GAN, the main advantages of VAE lie in its utilization of variational inference and reparameterization techniques, which make the optimization process smoother and more stable. More specifically, it characterizes the distribution of these hidden representations through an encoder-decoder learning paradigm and therefore uses the variance of the Gaussian distribution to describe the uncertainty of the input data well. In addition, the decoder maximizes the expected likelihood of input data conditioned on such latent variables, thereby reducing the deficiency of unexpected uncertainty. For instance, the highly representative ContrastVAE \cite{ContrastVAE} achieves impressive performance through variational modeling and contrastive learning. However, this method adopts data augmentation and model augmentation, which will destroy the semantic information of the original sequence, moreover, random augmentation operations will generate secondary contrastive views.

To resolve these issues, we put forward a novel Meta-optimized Seq2Seq Generator and Contrastive Learning model (Meta-SGCL) for sequential recommendation. 
Firstly, we leverage VAE as a Sequence-to-Sequence (Seq2Seq) generator to obtain augmentation views for contrastive learning. 
In this way, we can produce contrast views for input data generatively, without hand-crafted contrastive augmentation strategies, while keeping the semantic information of the input data. 
We further extend the traditional univariate Evidence Lower Bound (ELBO) of the VAE framework to two view cases for contrastive learning, and prove theoretically that optimizing double ELBO results in mutual information maximization terms in contrastive learning tasks.
Secondly, in order to enable our Seq2Seq generator to generate contrast views for different data adaptively, we further propose a training method based on a meta-optimized two-step training strategy for our framework without relying on manually designed view-generation techniques. Finally, experiments on three benchmark datasets to verify the effectiveness of the proposed model on the sequential recommendation task.

The main contributions are summarized as follows:
\begin{itemize}
\item To the best of our knowledge, this is the first work to use the generative method for producing contrastive views in sequential recommendation.
\item We propose a contrastive learning sequential recommendation model called Meta-SGCL to adaptively generate augmentation views by applying the meta-optimized two-step training strategy.
\item We conduct extensive experiments to demonstrate the effectiveness of our method, achieving state-of-the-art performance on three benchmark SR datasets.
\end{itemize}

% The rest of this paper is organized as follows. We first describe our problem statement in Section II. Then we provide the preliminaries in Section III. Next, we introduce our proposed model Meta-SGCL in Section IV, and present the experimental results in Section V. Section VI reviews the related work. Finally, Section VII concludes our paper.

\section{Problem Formulation}
We denote a user set and an item set as $\mathcal{U} =\left \{ u_1, u_2,..., u_M \right \} $ and $\mathcal{V} =\left \{ v_1,v_2,..., v_N \right \} $, where $M$ and $N$ represent the number of users and items, respectively.  $u\in \mathcal{U}$ represents a user and $v \in \mathcal{V}$ represents an item. In general, a user $u$ has a chronological sequence of interaction items $ s^u =[x_1, x_2,..., x_{n} ] $, where $x_{t} \in \mathcal{V}$ represents a clicked item of the user at time step $t$.  
The goal of sequence recommendation is to predict the $(n+1)$-th interaction item of a user based on the last $n$ interaction sequences. For each $s^u$, we further introduce a latent variable $z$ representing user's preference at time $n$.
The scores of all candidate items are denoted by $\hat{\boldsymbol{y}}=[\hat{y}_1, \hat{y}_2, ..., \hat{y}_{N}]$, where $\hat{y}_i$ refers to the score of item $v_i$. The prediction scores are ranked in descending order, and the items ranked in the top-$k$ are used for recommendation.

\section{Preliminaries}
\subsection{Variational AutoEncoder and Evidence Lower Bound}
Recently, Variational AutoEncoder (VAE) \cite{VAE} has been applied to many
issues. The VAE is a generative model which models variables as random distributions based on the Bayesian theorem. 
Assuming a variable $z$ being the sampled latent representation from sequence $s^u=[x_1,x_2,...,x_n]$, we aim to maximize the probability of the next item, that is, to maximize the probability of the whole sequence $s^u$:
\begin{equation}
\centering
\begin{aligned} 
p(s^u)=\prod\limits_{x_{n+1}^u\in s^u}p(x_{n+1}^u|x_1^u, x_2^u, \cdots,x_{n}^u)
\end{aligned}
\label{eq.1}
\end{equation}
The primary focus of the model is to effectively represent the joint probability $p(s^u)$.

In the VAE framework, we begin by assuming a continuous latent variable $z$, which is sampled from a standard normal distribution, denoted as $z \sim \mathcal{N}(0, I)$. This assumption is based on the intuition that variables following a standard normal distribution can capture intricate dependencies. 

A key characteristic of VAE is the use of highly flexible function approximators, such as neural networks, to parameterize the conditional distribution $p(s^u|z)$. This allows for modeling the potentially highly nonlinear mapping from the latent variable $z$ to the user's sequence $s^u$. In VAE, the joint probability $p(s^u|z)$ can be specified using the marginal distribution as follows:
\begin{equation}
p(s^u)=\int p(s^u|z)p(z)dz
\label{eq.2}
\end{equation}

Since the probability $p(s^u)$ is intractable, the variational inference method takes advantage of Bayesian Theorem $p(s^u,z)=p(s^u|z)p(z)$ and proposes a posterior distribution $q(z|s^u)$ to approximate the true distribution $p(z|s^u)$ \cite{DBLP:journals/corr/ZhaoSE17a}. Migrating to sequential recommendation, the log-likelihood of $p(s^u)$ can be derived as follows:
\begin{equation}
\begin {aligned}
\log p(S^u) &= \int q(z|s^u) \log p(s^u)  \mathrm{d}z \\
&\geq \mathbb{E}_{q}[\log p(s^u|z)]-D_{KL}[q(z|s^u)||p(z)]&
\end {aligned}
\label{eq.3}
\end{equation}
	
 Eq. (\ref{eq.3}) is the training objective of variational auto-encoder, it is called Evidence Lower BOund (ELBO).

As a result, the modeling needs the help of two neural networks: the encoder neural network infers the latent representation $z$ based on $s^u$ through $q(z|s^u$). The decoder neural network generates corresponding $s^u$ from the latent representation $z$ depending on $p(s_u|z)$. The learning process is controlled by the ELBO.

\subsection{Posterior collapse in VAE}
Despite the considerable success of VAE, they are often plagued by a significant issue known as posterior collapse, which severely limits the generative model's capacity \cite{DBLP:conf/nips/LucasTG019, DBLP:conf/aaai/ZhaoSE19}. Posterior collapse occurs when the KL divergence between the learned variational distribution $q(z|s^u)$ and the prior distribution $p(z)$ tends to zero per input $s^u$. This typically happens when the decoder model is too powerful, leading to the learned variational distribution becoming nearly identical to the prior, thereby making latent variables for different inputs indistinguishable in the latent space. Addressing the problem of posterior collapse has been the focus of recent research, with approaches attempting to mitigate the impact of the KL divergence term. Some methods down-weight the KL divergence term \cite{DBLP:conf/www/LiangKHJ18}, while others introduce an additional regularization term that explicitly maximizes the mutual information between the input and latent variables \cite{DBLP:journals/corr/abs-2102-08663}. 

However, in sequential recommendation tasks, the problem of posterior collapse becomes even more severe due to the sparse nature of user-item interactions and the difficulty in modeling users' dynamic preferences.
Despite these existing approaches, it has been found that they are insufficient to achieve significant performance improvements in sequential recommendation tasks. 
We propose to alleviate the above problems from a CL perspective from a contrastive learning perspective, which aims to maximize the mutual information between different views of the same sequence in the latent space $I(z,z')$.
In the Double ELBO section, it is shown that by extending the single latent variable generative model in the VAE framework to a two-view case, the VAE framework can naturally incorporate the principle of maximizing mutual information. This can be optimized using contrastive learning loss, which helps alleviate the problem of posterior collapse and improve performance in sequential recommendation tasks.

\section{Our Proposed Model}
\subsection{Overall Framework} 
\begin{figure*}[t]
	\centering
	\includegraphics[width=1\textwidth]{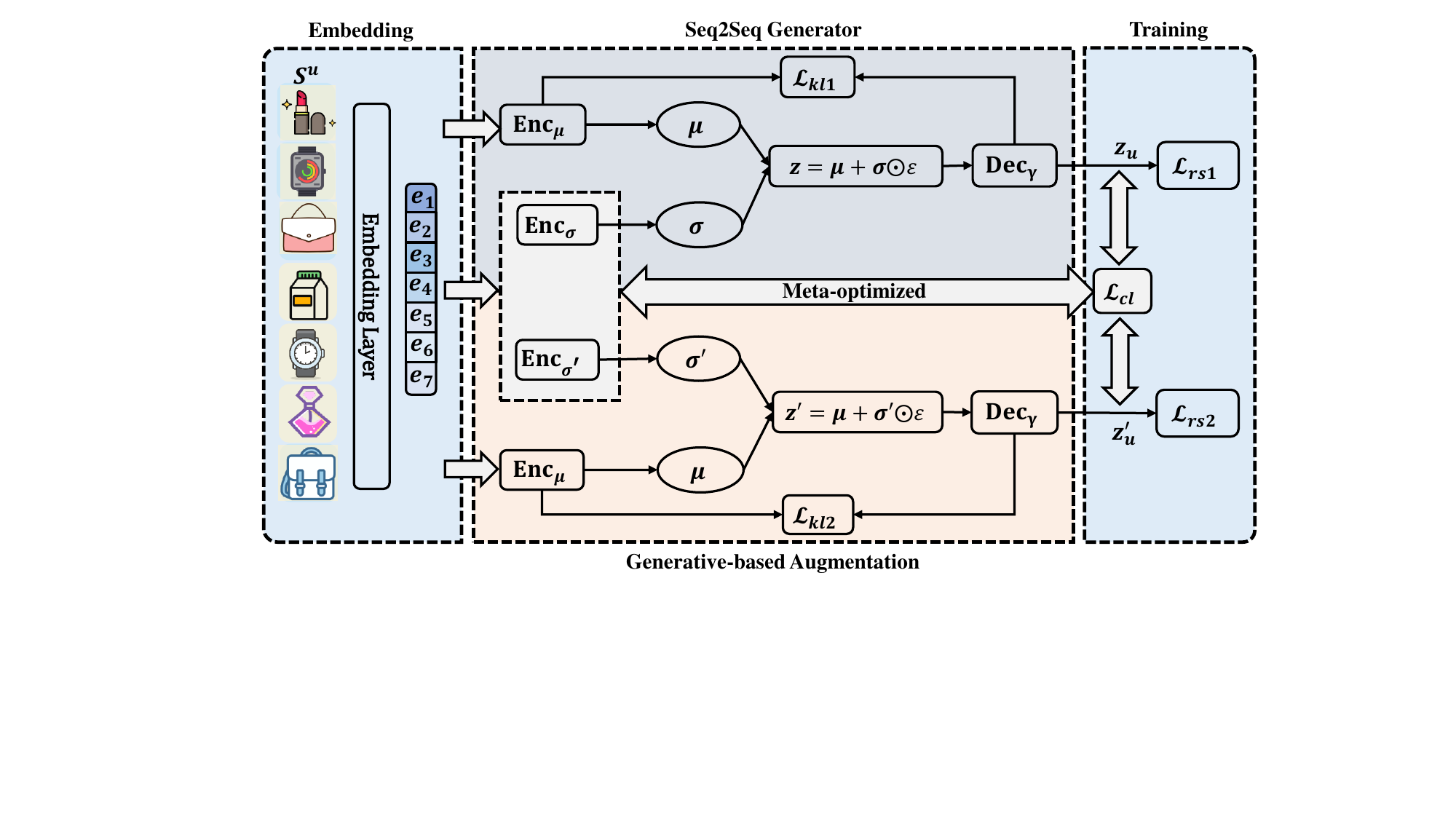}
    \vspace{-3.5cm}
	\caption{The overview of the proposed model Meta-SGCL. Meta-SGCL has the following main components: Embedding, Seq2Seq Generator ( Sequential Encoder, Sequential Decoder), Generative-based Augmentation, and Training.
	 } \label{fig2}
\end{figure*}

Figure \ref{fig2} shows the general framework of the Meta-optimized Seq2Seq Generator and Contrastive Learning (Meta-SGCL) model for sequential recommendation. Meta-SGCL has the following main components: 
1) Embedding layer aims to represent items as low-dimensional vectors. 2) Seq2Seq generator uses VAE as the backbone and leverages Transformer as the sequence encoder and decoder. It inputs the user's interaction sequence representation, and the output is the reconstructed user sequence, aiming to generate a new user sequence representation. 3) Generative-based augmentation aims to generate augmented views for contrastive learning. 4) Training proposes a double ELBO theorem and a new meta-optimized training method to adaptively generate augmentation views for contrastive learning. 

\subsection{Embedding Layer}
For the problem of transforming discrete variables such as item indexes into continuous vectors, we consider using the learnable item embedding table $\mathbf {M} \in \mathbb{R}^{N\times d}$, where $N$ is the total number of items and $d$ is the dimension in which the items are embedded. For any user interaction sequence, we first model all sequences as equal-length sequences $T$, for sequences larger than this value, we only keep items of the length of the most recent interaction; for sequences smaller than this length, we first padding with zeros makes all user sequences equal in length.
After that, we can represent all items in a matrix: $\mathbf{E} \in \mathbb{R} ^{n\times d}$, and the embedding of the item $v_t$ can be looked up from the table as $\mathbf {E}_i=\mathbf {M}_{v_i}$.

To describe the position relationship of the user, we embed the position information of the item, and the calculation method is as follows:

\begin{equation}
\hat{\mathbf{E}}=\mathbf{E}+\mathbf{P}=\begin{bmatrix} \mathbf{M}_{v_1}+\mathbf{P}_1
 \\ \mathbf{M}_{v_2}+\mathbf{P}_2
 \\...
 \\ \mathbf{M}_{v_n}+\mathbf{P}_T
\end{bmatrix}
\label{eq.4}
\end{equation}
where $\mathbf{M}_{v_i}$ is the item embedding of one user at time $t$ and $\mathbf{P}_t \in \mathbb{R} ^{n\times d}$ is the position embedding at position $t$.

\subsection{Seq2Seq Generator}
After getting the embedding of the sequence, the encoder-decoder framework is a widely used method to accomplish the Seq2Seq objective. The encoder encodes the input sequence into a context vector, which is then used by the decoder to generate the output sequence. 

\subsubsection{Sequential Encoder}
Inspired by the excellent performance of the Transformer method on the sequential recommendation, we use Transformer as the encoder of our model. 
 
The ultimate task of sequential recommendation is to recommend items that may be clicked next based on the user's historical interaction sequence. However, user interaction may have different importance for each item. The self-attention mechanism can adaptively learn weights to model the importance of users' access to each item so that the above problems can be solved. The representation of the self-attention layer is as follows:
\begin{equation}
    \mathbf{H}=softmax(\frac{(\mathbf{\hat{E}}\mathbf{W}^Q)(\mathbf{\hat{E}}\mathbf{W}^K)^\top}{\sqrt{d} } )(\mathbf{\hat{E}}\mathbf{W}^V)
    \label{eq.5}
\end{equation}
where $\mathbf{W}^Q$, $\mathbf{W}^K$ $\mathbf{W}^V \in \mathbb{R} ^{d\times d}$ are learnable weight parameters whose values are learned by the softmax function. $\sqrt{d}$ can effectively avoid a large normalization factor in the softmax function. In order to make the recommendation results more reliable, when we learn item representation, we block all items after the current moment to avoid information leakage.

To facilitate the model's ability to simultaneously consider information from various representation subspaces at different positions, we employ multi-head attention, utilizing $h$ separate attention models running in parallel, each with distinct parameters. The outputs of all attention models are concatenated to generate the final values.
\begin{equation}
    \mathbf{O}=Concat(\mathbf{H}_1, \mathbf{H}_2, ..., \mathbf{H}_h)
    \label{eq.6}
\end{equation}
\begin{equation}
    \mathbf{H}_i=softmax(\frac{(\mathbf{\hat{E}}\mathbf{W}_i^Q)(\mathbf{\hat{E}}\mathbf{W}_i^K)^\top}{\sqrt{d} } )(\mathbf{\hat{E}}\mathbf{W}_i^V)
    \label{eq.7}
\end{equation}
where the projection matrices $\mathbf{W}_i^Q$, $\mathbf{W}_i^K$, $\mathbf{W}_i^V \in \mathbb{R} ^{d\times d/h}$

The self-attention mechanism operates linearly, and at the output layer, we use a multi-layer perceptron as output. The nonlinear activation function is ReLU, and then we add operations such as residual connection, normalization, and random dropout to our model. Finally, multiple self-attention layers make up our encoder. 
\begin{equation}
\mathbf{F}=\mathrm{ReLU}(\mathbf{O}\mathbf{W}_1+\mathbf{b}_1)\mathbf{W}_2+\mathbf{b}_2+\mathbf{O}
\label{eq.8}
\end{equation}
where $\mathbf{W}_1, \mathbf{W}_2$ are $d \times d$ matrices and $\mathbf{b}_1, \mathbf{b}_2 \in \mathbb{R}^{d}$ are bias vectors. We also employ normalization to normalize the inputs across features while dropout is used to avoid overfitting.

In order to capture more intricate item transitions, we employ a stacked self-attention block to build a multi-layer self-attention structure. To simplify the explanation, we can define the complete SAN as follows:
\begin{equation}
\mathbf{F}=\mathrm{SAN}(\mathbf{E})
\label{eq.9}
\end{equation}
Then the $l$-th $(l>1)$ SAN layer is defined as:
\begin{equation}
\mathbf{F}^{(l)}=\mathrm{SAN}(\mathbf{F}^{(l-1)})
\label{eq.10}
\end{equation}
where $\mathbf{F}^{(l)}$ is the final output of the multi-layer SAN.

We then take $\mathbf{F}_i^{(l)}$ as the input of the VAE framework. Let $\boldsymbol{z}$ be the latent variable sampled from the sequence ${s}^u$, which follows a Gaussian distribution. We infer the posterior distribution $q(\boldsymbol{z}|{s}^u)$ as a multinomial layer. The mean and variance vectors are computed based on the self-attention vectors:
\begin{equation}
\boldsymbol{\mu}=Enc_{{\mu}}(\mathbf{F}_i^{(l)}), \ \boldsymbol{\sigma}=Enc_{{\sigma}}(\mathbf{F}_i^{(l)})
\label{eq.11}
\end{equation}
where $Enc_{\mu}(\cdot)$ and $Enc_{\sigma}(\cdot)$ represent linear transformations. 

By using the reparameterization trick, the output of our sequential encoder is written as:
\begin{equation}
\boldsymbol{z}=\boldsymbol{\mu}+\boldsymbol{\sigma}\odot\epsilon
\label{eq.12}
\end{equation}
where $\epsilon \sim \mathcal{N}(0,\mathrm{I})$. By sampling a random variable $\epsilon$ with standard Gaussian distribution, the latent representation of the sequence is reparameterized.

\subsubsection{Sequential Decoder}
The decoder only learns the potential generation process of the data according to the estimated distribution of the latent factors and recommends the next possible interactive item to the user by maximizing the reconstruction likelihood. The process is as follows:
\begin{equation}
\begin{aligned}
\mathbb{E}_{q(\boldsymbol{z}|{s}^u)}\mathrm{log} [p({s}^u|\boldsymbol{z})] 
=\mathbb{E}_{q(\boldsymbol{z}|{s}^u)}\mathrm{log} [({s}^u|\boldsymbol{z})] 
\label{eq.13}
\end{aligned}
\end{equation}

In Meta-SGCL, the decoder $Dec$ is chosen as the same Transformer architecture as the encoder.

\subsection{Generative-based Augmentation}
To obtain an augmentation view for contrastive learning, we design a generative-based augmentation 
operator to create the augmentation view for each user. Figure \ref{fig2} illustrates the structure of the generative-based augmentation 
operator for contrastive learning. By feeding the same sequence into the Seq2Seq generator again, we can get different standard deviations. And we multiply different standard deviations by the Gaussian noise to obtain different views When using the reparameterization technique. These two samples can be mutually augmentation views and be used for contrastive learning. In this way, we generate augmentation views largely without distorting the semantic information and sequential patterns of user interaction sequences.

Taking $\mathbf{F}_i^{(l)}$ as the input, we get the mean and another variance vector $\boldsymbol{\sigma}'$ as:
\begin{equation}
\boldsymbol{\mu}=Enc_{{\mu}}(\mathbf{F}_i^{(l)}), \ \boldsymbol{\sigma}'=Enc_{{\sigma}'}(\mathbf{F}_i^{(l)})
\label{eq.14}
\end{equation}

 By using the reparameterization trick again, the augmentation view is obtained by the following formula:
\begin{equation}
\boldsymbol{z}'=\boldsymbol{\mu}+{\boldsymbol{\sigma}'}\cdot \epsilon
\label{eq.15}
\end{equation}

\subsection{Model Training}
In this section, we introduce the double ELBO theorem and the meta-optimized update strategy used in Meta-SGCL model, and explain how to adaptively generate contrastive views for contrastive learning.

\subsubsection{Double ELBO} Consider a generative model with one observed sequence $s^u$ and two latent variables $\boldsymbol{z}$ and $\boldsymbol{z}'$ with structure $s^u\gets \boldsymbol{z} - \boldsymbol{z}'\to s^u$, where $\boldsymbol{z}$ and $\boldsymbol{z}'$ are used to generate $s^u$. $\boldsymbol{z}$ and $\boldsymbol{z}'$ are dependent on the same input. Then we have the following double lower bound of the log joint probability of the observed variables:
\begin{equation}
\begin{aligned}
&\log p(s^u,\ s^u)\\
&\ge \mathbb{E}_{q}[\log p(s^u|\boldsymbol{z})]-D_{KL}[q(\boldsymbol{z}|s^u)||p(\boldsymbol{z})] \\
&+\mathbb{E}_{q}[\log p(s^u|\boldsymbol{z}')]-D_{KL}[q(\boldsymbol{z}'|s^u)||p(\boldsymbol{z}')] \\
&+\mathbb{E}_{q(\boldsymbol{z},\boldsymbol{z}'|S^u, S^u)}\log\left[\frac{p(\boldsymbol{z},\boldsymbol{z}')}{p(\boldsymbol{z})p(\boldsymbol{z}')}\right]&
\end{aligned}
\label{eq.16}
\end{equation}

According to the above generative model, we have $s^u$ that are conditionally independent give $\boldsymbol{z}$ and $\boldsymbol{z}'$, or formally $p(s^u, s^u|\boldsymbol{z},\boldsymbol{z}')$, then we can approximate the posterior with a variational distribution $q(\boldsymbol{z},\boldsymbol{z}'|s^u, s^u)$ which could be factorized through:
\begin{equation}
q(\boldsymbol{z},\boldsymbol{z}'|s^u, s^u)=q(\boldsymbol{z}|s^u)q(\boldsymbol{z}'|s^u)
\label{eq.17}
\end{equation}

Then we have:
\begin{equation}
\begin{aligned}
&\log p(s^u,\ s^u)=\log\int p(s^u, s^u,\boldsymbol{z},\boldsymbol{z}')d\boldsymbol{z}d\boldsymbol{z}' \\
&=\log\mathbb{E}_{q(\boldsymbol{z},\boldsymbol{z}'|s^u, s^u)}\left[\dfrac{p(s^u, s^u,\boldsymbol{z},\boldsymbol{z}')}{q(\boldsymbol{z},\boldsymbol{z}'\lvert s_u,s_u)}\right] \\
&\geq\mathbb{E}_{q(\boldsymbol{z},\boldsymbol{z}'\mid s^u, s^u)}\log\left[\dfrac{p(s^u, s^u,\boldsymbol{z},\boldsymbol{z}')}{q(\boldsymbol{z},\boldsymbol{z}'|s^u, s^u)}\right] \\
&=\mathbb{E}_{q(\boldsymbol{z},\boldsymbol{z}'|s^u, s^u)}\log\left[\dfrac{p(s^u|\boldsymbol{z})p(s^u|\boldsymbol{z}')p(\boldsymbol{z},\boldsymbol{z}')}{q(\boldsymbol{z}|s^u)q(\boldsymbol{z}'|s^u)}\right] \\
&=\mathbb E_{q(\boldsymbol{z}|s^u)}\log[p(s^u|\boldsymbol{z})] + \mathbb E_{q(\boldsymbol{z}'|s^u)}\log[p(s^u|\boldsymbol{z}')] \\
&+\mathbb{E}_{q(\boldsymbol{z},\boldsymbol{z}'|s^u,s^u)}\log\left[\dfrac{p(\boldsymbol{z},\boldsymbol{z}')}{q(\boldsymbol{z}|s^u)q(\boldsymbol{z}'|s^u)}\right]&
\end{aligned}
\label{eq.18}
\end{equation}
The last term in the last equation could be further expanded:
\begin{equation}
\begin{aligned}
&\mathbb{E}_{q(\boldsymbol{z},\boldsymbol{z}'|s^u,s^u)}\log\left[\dfrac{p(\boldsymbol{z},\boldsymbol{z}')}{q(\boldsymbol{z}|s^u)q(\boldsymbol{z}'|s^u)}\right] \\
&=\mathbb{E}_{q(\boldsymbol{z},\boldsymbol{z}'\vert s^u,s^u)}\log\left[\dfrac{p(\boldsymbol{z},\boldsymbol{z}')p(\boldsymbol{z})p(\boldsymbol{z}')}{q(\boldsymbol{z}\vert s^u)q(\boldsymbol{z}'\vert s^u)p(\boldsymbol{z})\it p(\boldsymbol{z}')}\right]\quad \\
&=\mathbb{E}_{q(\boldsymbol{z},\boldsymbol{z}'|S^u,s^u)}\log\left[\dfrac{p(\boldsymbol{z},\boldsymbol{z}')}{p(\boldsymbol{z})p(\boldsymbol{z}')}\right] \\
&+\mathbb{E}_{q(\boldsymbol{z},\boldsymbol{z}'|s^u,s^u)}\log\left[\dfrac{p(\boldsymbol{z})p(\boldsymbol{z}')}{q(\boldsymbol{z}|s^u)q(\boldsymbol{z}'|s^u)}\right]\quad \\
&=\mathbb{E}_{q(\boldsymbol{z},\boldsymbol{z}'|s^u,s^u)}\log\left[\frac{p(\boldsymbol{z},\boldsymbol{z}')}{p(\boldsymbol{z})p(\boldsymbol{z}')}\right]\\ 
&-D_{KL}[q(\boldsymbol{z}|s^u)||p(\boldsymbol{z})]-D_{KL}\left[q(\boldsymbol{z}'|s^u)||p(\boldsymbol{z}')\right]
\end{aligned}
\label{eq.19}
\end{equation}

Plugging Eq. (\ref{eq.19}) into Eq. (\ref{eq.18}), then we can obtain Eq. (\ref{eq.16}).

Note that the first four terms on the right of Eq. (\ref{eq.16}) are identical to that of the vanilla ELBO in Eq. 
(\ref{eq.3}) and could be effectively optimized using traditional VAE models. The last term $\mathbb{E}_{q(\boldsymbol{z},\boldsymbol{z}'|s^u, s^u)}\log\left[\frac{p(\boldsymbol{z},\boldsymbol{z}')}{p(z)p(z')}\right]$, however, is hard to compute. To make this term tractable, we follow the practice in aitchison that specifies $p(\boldsymbol{z},\boldsymbol{z}')=q(\boldsymbol{z},\boldsymbol{z}')$, $p(\boldsymbol{z})=q(\boldsymbol{z})$ and $p(\boldsymbol{z}')=q(\boldsymbol{z}')$ through choosing specific prior distributions, and then this term becomes $\mathbb{E}_{q(\boldsymbol{z},\boldsymbol{\boldsymbol{z}}'|s^u, s^u)}\log\left[\frac{q(\boldsymbol{z},\boldsymbol{z}')}{q(\boldsymbol{z})q(\boldsymbol{z}')}\right]$. If taking its expectation under the true data distribution $p(s^u,s^u)$, the last term becomes:
\begin{equation}
\begin{aligned}
\mathbb{E}_{q(\boldsymbol{z},\boldsymbol{z}^{\prime})}\frac{q(\boldsymbol{z},\boldsymbol{z}^{\prime})}{q(\boldsymbol{z})q(\boldsymbol{z}^{\prime})} 
&=D_{KL}[q(\boldsymbol{z},\boldsymbol{z}^{\prime})||q(\boldsymbol{z})q(\boldsymbol{z}^{\prime})]\\
&=I(\boldsymbol{z},\boldsymbol{z}^{\prime})
\label{eq.20}
\end{aligned}
\end{equation}
Eq. (\ref{eq.20}) indicates that we can maximize the mutual information between $\boldsymbol{z}$ and $\boldsymbol{z}'$ from $q(\boldsymbol{z},\boldsymbol{z}')$. Note that $q(\boldsymbol{z},\boldsymbol{z}')$ are the encoder’s output taking $s^u$ twice as input, so the mutual information term can be efficiently estimated using its tractable lower bounds through CL.

\subsubsection{Meta-optimized Training Strategy}
In this section, we introduce a novel meta-optimized update strategy that enables our model to adaptively generate contrastive views for downstream tasks. To optimize the reconstruction term, we formalize it as a next-item recommendation task where the log-likelihood could be factorized
as follows:
\begin{equation}
\mathrm {log}p({s}^u|\boldsymbol{z})\propto \hat{\mathbf{y}} \   \mathrm {and} \ \hat{\mathbf{y}}'
\label{eq.21}
\end{equation}

The sequential recommendation  framework is shown in Figure \ref{fig2}. 
The sequence representation is $\boldsymbol{z}_u$, and the item embedding matrix is $\mathbf{M}$, then the recommendation score of the item is calculated as follows:
\begin{equation}
\begin{aligned}
\hat{\mathbf{y}}=\boldsymbol{z}_u{\mathbf{M}^{\top}},
\\ \hat{\mathbf{y}}'=\boldsymbol{z}'_u{\mathbf{M}^{\top}}
\end{aligned}
\label{eq.22}
\end{equation}
where $\hat{\mathbf{y}}, \hat{\mathbf{y}}' \in \mathbb{R}^{\mathcal{V}}$. 
After converting the index of the ground truth item into a one-hot vector $\mathbf{y}$, we use the cross-entropy loss function to calculate the score of each item as follows:
\begin{equation}
\begin{aligned}
\mathcal{L}_{rs_1} = -\mathrm{one\mbox{-}hot} (\mathbf{y}) \mathrm {log} (\hat{\mathbf{y}}), \\ 
\mathcal{L}_{rs_2} = -\mathrm{one\mbox{-}hot} (\mathbf{y}) \mathrm {log} (\hat{\mathbf{y}}'),
\end{aligned}
\label{eq.23}
\end{equation}

Note that the KL divergence term for $({s}^u,\boldsymbol{z})$ are identical and we illustrate that for $({s}^u,\boldsymbol{z})$. The KL-divergence term can be easily computed as follows:
\begin{equation}
\begin{aligned}
\mathcal{L}_{kl_1}
&=D_{KL_1}[q(\boldsymbol{z}|{s}^u)||p(\boldsymbol{z})] \\
&= {{\textstyle \sum_{u=1}^{M}}({(\boldsymbol{\sigma})}^2+{(\boldsymbol{\mu})}^2-1-\mathrm {log} {{(\boldsymbol{\sigma})}^2)}}
\end{aligned}
\label{eq.24}
\end{equation}
In the same way, the optimizations of the reconstruction term and KL divergence term for $({s}^u,\boldsymbol{z}')$ are identical and we illustrate that for $({s}^u,\boldsymbol{z}')$. The KL-divergence term can be easily computed as follows:
\begin{equation}
\begin{aligned}
\mathcal{L}_{kl_2}
&=D_{KL_2}[q(\boldsymbol{z'}|{s}^u)||p(\boldsymbol{z'})] \\
&= {{\textstyle \sum_{u=1}^{M}}({(\boldsymbol{\sigma'})}^2+{(\boldsymbol{\mu})}^2-1-\mathrm {log} {{(\boldsymbol{\sigma'})}^2)}}
\end{aligned}
\label{eq.25}
\end{equation}
where $M$ is the number of users.

To maximize the mutual information term between $\boldsymbol{z}$ and $\boldsymbol{z'}$ under $q(\boldsymbol{z}, \boldsymbol{z'})$, we adopt the InfoNCE loss function, which is a multi-sample unnormalized lower bound of mutual information with low variance. Denote $u$’s hidden representation by $\boldsymbol{z_u}$, and its positive sample by $\boldsymbol{z_u'}$, we choose InfoNCE as the loss function for contrastive learning to maximize the mutual information of $u$'s two representations. The contrastive learning loss function between them is calculated as follows:
\begin{equation}
\begin{aligned}
\mathcal{L}_{cl}
&=\mathcal{L}_{InfoNCE} \\
&=\frac{1}{M}{\sum_{u=1}^{M}}\mathrm{log}\frac{\mathrm {exp}(\boldsymbol{z}_u^{\top}, \boldsymbol{z}_u'/\tau)}{\mathrm {exp}(\boldsymbol{z}_u^{\top},\boldsymbol{z}_u'/\tau)+{\textstyle \sum_{v\ne u}} \mathrm {exp}(\boldsymbol{z}_u^{\top},\boldsymbol{z}_v/\tau)}
\end{aligned}
\label{eq.26}
\end{equation}
where $\tau$ is the temperature parameter. Intuitively, the loss function requires that the similarity between the $m$-th sample and its positive sample is as large as possible, and the similarity with negative samples is as large as possible small. $\boldsymbol{z}_u$ is a representation of the $u$-th sequence, which summarizes the latent representations of all tokens of the sequence. We regard $\boldsymbol{z}_u'$ as the positive sample and $\boldsymbol{z}_v$ is the negative sample.

Plugging Eqs. (\ref{eq.23})-(\ref{eq.26}) into the double ELBO in Eq. (\ref{eq.16}), we get the final objective function:
\begin{equation}
\mathcal{L} =(\mathcal{L}_{rs1}-\beta\mathcal{L}_{kl1})+(\mathcal{L}_{rs2}-\beta\mathcal{L}_{kl2})+\alpha{\mathcal{L}_{cl}},
\label{eq.27}
\end{equation}

Let $\mathcal{L}_{rs}=(\mathcal{L}_{rs1}+\mathcal{L}_{rs2})$  and $\mathcal{L}_{kl}=(\mathcal{L}_{kl1}+\mathcal{L}_{kl2})$, after simplifying the above formula we can get:
\begin{equation}
\begin{aligned}
&\mathcal{L} =(\mathcal{L}_{rs1}+\mathcal{L}_{rs2})-(\beta\mathcal{L}_{kl1}+\beta\mathcal{L}_{kl2})+\alpha{\mathcal{L}_{cl}}\\
&=\mathcal{L}_{rs}+\alpha \mathcal{L}_{cl}-\beta \mathcal{L}_{kl}
\label{eq.28}
\end{aligned}
\end{equation}
where $\alpha$ and $\beta$ are hyper-parameters that need to be tuned. Inspired by KL annealing, we introduce a $\beta$ to control the KL term on the above loss function. KL annealing is a common heuristic method used to train VAEs when the model is not fully utilized. We only need to multiply the KL term by a weight coefficient, which is $\beta$ in our work.

% \begin{algorithm}[t]
% \caption{Meta-SGCL algorithm}
% \label{alg:algorithm}
% \textbf{Input}:Training dataset ${\{S_u\}}^{N}_{u=1}$; Learning rate $l$ and $l^{'}$; Hyper-parameters $\alpha, \beta$ and $\tau$.\\
% \textbf{Parameter}: $\mu$ for $\mathrm{Enc}_{\mu}$; $\sigma_1$ and $\sigma_2$ for $\mathrm{Enc}_{\sigma_1}$ and  $\mathrm{Enc}_{\sigma_2}$; $\phi$ for $\mathrm{Dec}.$
%  \begin{algorithmic}[1] %[1] enables line numbers
%         \FOR{$t$-th training iteration}
%         \STATE $\mathcal{L} =\mathcal{L}_{rs}+\alpha \mathcal{L}_{cl}+\beta \mathcal{L}_{kl}$
%         \STATE Update $\mathrm{Enc}_{\sigma_1}$, $\mathrm{Enc}_{\mu}$ and $\mathrm{Dec}$ by: 
%         \STATE $\sigma_1,\mu,\phi \leftarrow \sigma_1,\mu,\phi - l \cdot \nabla_{\sigma_1,\mu,\phi}\mathcal{L}$
%         %\STATE \textcolor{gray}{\#comment}
%         \ENDFOR
%         \FOR{$t$-th training iteration}
%         \STATE $\mathcal{L'} =\mathcal{L}_{cl}$
%         \STATE Update $\mathrm{Enc}_{\sigma_2}$ by:
%         \STATE $\sigma_2 \leftarrow \sigma - l^{'} \cdot \nabla_{\sigma_1}\mathcal{L}^{'}$
%         \ENDFOR
%     \end{algorithmic}
% \end{algorithm}
We perform a meta-optimized strategy to guide the training the model, which is beneficial for the model to mine discriminative augmentation views from the sequence. The whole training process can be concluded in two stages. 

Specially, in the first stage, we calculate the sequential recommendation loss, KL divergence loss and contrastive loss to update the $Enc_{\boldsymbol{\sigma}}$, $Enc_{\boldsymbol{\mu}}$ and $Dec_{\boldsymbol{\gamma}}$ by back-propagation, which can be calculated via Eq. (\ref{eq.28}). In the second stage, we fix the parameters of $Enc_{\boldsymbol{\sigma}}$, $Enc_{\boldsymbol{\mu}}$ and and $Dec_{\boldsymbol{\gamma}}$ and obtain the temporary meta encoder $Enc_{\boldsymbol{\sigma}'}$ according the performance of the original encoder. Denote $\boldsymbol{\sigma}$ is the learned parameters by back-propagation at the first stage, we use the learned $\boldsymbol{\sigma}$ to re-encode the sequence by Eq. (\ref{eq.28}), recompute $\mathcal{L'}$ by Eq. (\ref{eq.26}), and then leverage back-propagation to obtain the parameters of the meta encoder $Enc_{\boldsymbol{\sigma}'}$. 
% The loss is calculated as follows:
% \begin{equation}
% \mathcal{L'} =\mathcal{L}_{cl}
% \label{eq.29}
% \end{equation}
% The encoders and decoders are iteratively trained until convergence. The details of our proposed meta-training strategy are presented in \textbf{Algorithm} \ref{alg:algorithm}

% \begin{table}[t]
% \renewcommand{\arraystretch}{1.15}
% \centering
% \resizebox{0.48\textwidth}{!}{
% \begin{tabular}{c|ccccc}
% \hline
% Dataset  & users    & items    & interactions  & avg.length  & sparsity \\ \hline
% Clothing & 39,387   & 23,033   & 278,677       & 7.1         & 99.97$\%$ \\ 
% Toys     & 19,412   & 11,924   & 167,597       & 8.6         & 99.93$\%$ \\ 
% ML-1M    & 6040     & 3416     & 999,611       & 165.5       & 95.16$\%$  \\ 
% \hline
% \end{tabular}}
% \caption{Datasets statistics.}
% \label{datasets}
% \end{table}

\subsection{Complexity Analysis}
\subsubsection{Time Complexity} Our model's time consumption primarily arises from its encoder and decoder, each exhibiting a time complexity of $O(n^2d+nd^2)$. The main contributors to this complexity are the self-attention network layer and the feed-forward network layer. Notably, the self-attention network's primary cost is $O(n^2d)$, while the feedforward network layer's main cost is $O(nd^2)$. While traditional RNNs boast a time complexity of $O(nd^2)$, CNNs operate at $O(wnd^2)$, where $w$ signifies the window size of the convolutional filter. In practice, our model significantly outperforms RNNs and CNNs in terms of speed. The efficiency of our model is further enhanced by the fully parallelizable computations within each self-attention layer, making it highly suitable for GPU acceleration. Consequently, training times are substantially reduced, solidifying the practical value of our time-efficient approach.

\subsubsection{Space Complexity} 
The predominant space consumption in our model can be attributed to the parameters utilized in item embedding, encoder layer, and decoder layer, which encompass crucial components such as self-attention calculation, feed-forward network, and layer normalization. Consequently, the overall space complexity of the Meta-SGCL model can be expressed as $O(Nd+nd+d^2)$.

\section{Experiments}
In this section, we first briefly present the datasets, baseline methods, evaluation metrics, and our implementation details in experimental settings. Then, we compare our proposed model Meta-SGCL with baseline methods, show the experimental results of all models and analyze the reasons. Moreover, we study the influence of model components on the performance of our model Meta-SGCL. Finally, we discuss the impact of some key parameters on the model results. Specifically, to study the validity of our model Meta-SGCL, we conduct rich experiments on three datasets to answer the following research questions (\textbf{RQ}s):
\begin{itemize}
\item  \textbf{RQ1}: Does our Meta-SGCL model perform compared to existing state-of-the-art baselines?
\item  \textbf{RQ2}: How does meta-optimized training strategy affect recommendation performance?
\item  \textbf{RQ3}: Is each component in the Meta-SGCL model necessary to improve the performance of the model?
\item  \textbf{RQ4}: How do key hyper-parameters in Meta-SGCL models affect model performance?
\item  \textbf{RQ5}: How does the robustness of Meta-SGCL?
\item  \textbf{RQ6}: How the contrastive regularization affects the training?
\end{itemize}

\subsection{Setups}
\textbf{Dataset.} Amazon\footnote{http://jmcauley.ucsd.edu/data/amazon/links.html} and MovieLens\footnote{https://grouplens.org/datasets/movielens/1m/} datasets are widely used in sequential recommendation. Moreover, the datasets belong to different domains and have different sparsity, making experimental results more universal and convincing.  For the Amazon dataset, as an E-commerce platform dataset, we apply the Clothing Shoes and Jewelry (Clothing) and Toys and Games (Toys) based on a 5-core version and filter out users who have interacted with less than five items. We binarize explicit data by discarding ratings of less than four. For MovieLens, we adopt the version MovieLens-1M (ML-1M) and perform the same operations as the Amazon dataset. For each user, we use the last clicked item for testing, the penultimate one for validation, and the remaining clicked items for training.  
The detailed information of the public datasets are summarizes in Table \ref{datasets}.

\textbf{Metrics.} For each user, we calculate the correlation score between the user and all users, and select the $k$ items with the highest scores for recommendation. The evaluation indicators are the most commonly used Hits Ratio (HR) and Normalized Discounted Cumulative Gain (NDCG). The larger the values of three metrics, the better the performance is. In the experimental results we report, the values of $k$ are 5 and 10.

\textbf{Implementation Details.} Our model is implemented in PyTorch. We adopt the Adam optimizer as the optimizer. Our model and all baselines are implemented on an Nvidia V100 GPU with 32G memory. To avoid overfitting, we use an early stopping strategy that stops running when there is no improvement within 100 epochs. The embedding dimension in the model is 64, the learning rate is 0.001, the number of attention heads is 2 and the dropout rate is 0.2. Moreover, we set the maximum sequence length $n$ as 200 on the ML-1M dataset and 50 on the Amazon datasets respectively. We conducted multiple experiments to ensure that the error of every experimental result is negligible. 

\begin{table}[t]
\renewcommand{\arraystretch}{1.2}
\centering
\caption{Datasets statistics.}
\resizebox{0.36\textwidth}{!}{
\begin{tabular}{c|ccc}
\hline
Dataset         & Clothing          & Toys           & ML-1M \\ \hline
users           & 39,387            & 19,412         & 6,040   \\ 
items           & 23,033            & 11,924         & 3,416    \\ 
interactions    & 278,677           & 167,597        & 999,611  \\ 
avg.length      & 7.1               & 8.6            & 165.5     \\
sparsity        & 99.97$\%$         & 99.93$\%$      & 95.16$\%$  \\
\hline
\end{tabular}}
\label{datasets}
\end{table}

\begin{table*}[t]
\renewcommand\arraystretch{1.22}
\centering
\caption{Performance comparisons of different methods. Where the bold score is the best in each row and the second-best baseline is underlined. The last column is the relative improvements compared with the best baseline results.}
\resizebox{0.9\textwidth}{!}{
\begin{tabular}{c|l|cc|ccccc|ccc|c|c}
\hline
Dataset 
& Metric 
& Pop & BPR-MF 
& GRU4Rec & Caser  & SASRec & BERT4Rec  & VSAN
& ACVAE   & DuoRec    & ContrastVAE 
& Meta-SGCL 
& Impro. \\ \hline
\multirow{4}{*}{Clothing} 
&HR@5     
&0.0042  &0.0067&0.0095  &0.0108  &0.0168  &0.0125   &0.0152
&0.0164   &\underline{0.0193}  &0.0159
&\textbf{0.0216}
&11.92$\%$ \\
&HR@10
&0.0076  &0.0094
&0.0165  &0.0174  &0.0272  &0.0208   &0.0246
&0.0255   &\underline{0.0302}  &0.0283
&\textbf{0.0309}
&2.32$\%$ \\
&NDCG@5
&0.0032  &0.0052
&0.0061  &0.0067  &0.0091  &0.0075   &0.0090
&0.0098  &\underline{0.0113}   &0.0102
&\textbf{0.0142}
&25.66$\%$ \\
&NDCG@10
&0.0045  &0.0069
&0.0083  &0.0098  &0.0124  &0.0102   &0.0106
&0.0120   &\underline{0.0148}  &0.0135
&\textbf{0.0167}
&12.84$\%$ \\
\hline
\multirow{4}{*}{Toys} 
&HR@5
&0.0065  &0.0120
&0.0121  &0.0205  &0.0429  &0.0371  &0.0472
&0.0457   &0.0539   &\underline{0.0548}
&\textbf{0.0642}
&17.15$\%$ \\
&HR@10
&0.0090  &0.0179
&0.0184  &0.0333  &0.0652  &0.0524  &0.0689
&0.0663    &0.0744  &\underline{0.0760}
&\textbf{0.0907}
&19.34$\%$ \\
&NDCG@5
&0.0044  &0.0067
&0.0077  &0.0125  &0.0248  &0.0259 &0.0328
&0.0291   &0.0340    &\underline{0.0353}
&\textbf{0.0420}
&18.98$\%$ \\
&NDCG@10
&0.0052  &0.0090
&0.0097  &0.0168  &0.0320  &0.0309   &0.0395
&0.0364   &0.0406  &\underline{0.0441}
&\textbf{0.0506}
&14.74$\%$ \\
\hline
\multirow{4}{*}{ML-1M} 
&HR@5
&0.0078  &0.0068
&0.0763  &0.0816  &0.1087  &0.0733   &0.1210
&0.1356  &\underline{0.2038}   &0.1152
&\textbf{0.2387}
&17.12$\%$ \\
&HR@10
&0.0162  &0.0162
&0.1658  &0.1593  &0.1904  &0.1323   &0.1815
&0.2033   &\underline{0.2946}  &0.1894
&\textbf{0.3560}
&20.84$\%$ \\
&NDCG@5
&0.0052  &0.0052
&0.0385  &0.0372  &0.0638  &0.0432   &0.0634
&0.0837  &\underline{0.1390}   &0.0687
&\textbf{0.1622}
&16.69$\%$ \\
&NDCG@10
&0.0079  &0.0079
&0.0671  &0.0624  &0.0910  &0.0619   &0.0881
&0.1145   &\underline{0.1680}  &0.0935
&\textbf{0.1953}
&16.25$\%$ \\
\hline
\end{tabular}}
\label{Overall Comparison}
\end{table*}

\begin{figure*}[ht]
	\centering
	\includegraphics[width=0.9\textwidth]{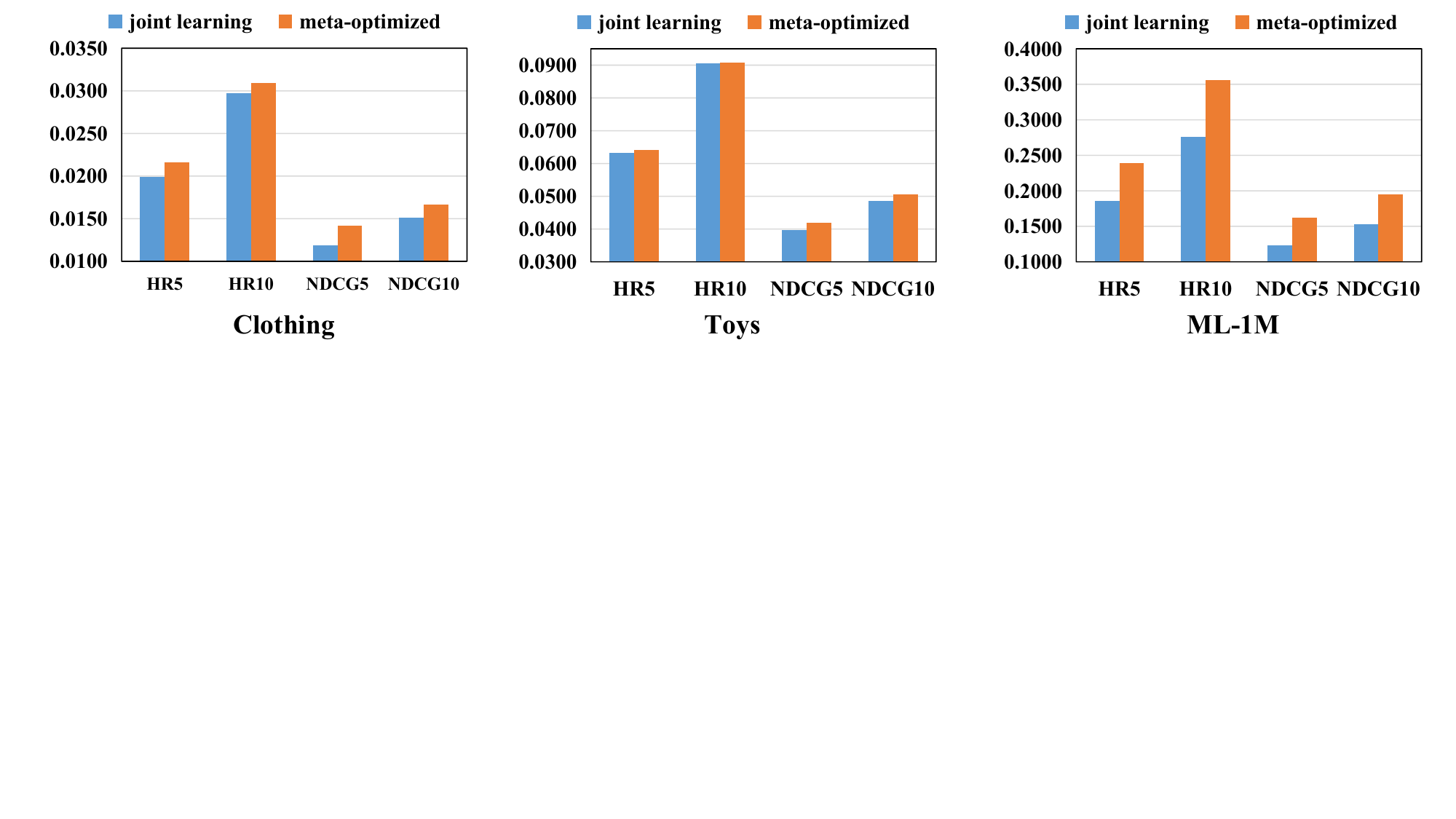}
        \vspace{-5.5cm}
	\caption{The performance of joint learning and meta-optimized two-step training on three datasets.
	 } \label{2S}
\end{figure*}

\textbf{Baselines.} We compare our methods with three types of representative SR models. Now, in order to better understand these models, we briefly introduce our competitors: 

\textbf{\textit{1) Traditional Recommendation Methods:} }
\begin{itemize}
 \item \textbf{Pop} is a non-personalized approach which recommends the same items for each user. These items are the most popular items which have the largest number of interactions in the whole item set.
 \item \textbf{BPR-MF} \cite{DBLP:conf/uai/RendleFGS09} designs a pair-wise optimization method combining with the matrix factorization model. This is the classic method of constructing recommendation from implicit feedback data.
\end{itemize}

\textbf{\textit{2) Sequential Recommendation Methods:} }
\begin{itemize}
\item \textbf{GRU4Rec} \cite{srnn2016} contains GRUs to model user interaction sequences and utilizes session-parallel mini-batches as well as a pair-wise loss function for training, which is the first RNN-based model for sequential recommendation.
\item \textbf{Caser} \cite{tang2018caser} is a CNN-based approach that captures high-order patterns for sequential recommendation by applying horizontal and vertical convolution operations.
\item \textbf{SASRec} \cite{DBLP:conf/icdm/KangM18}  leverages self-attention networks for sequential recommendation. It can not only model long-term preferences but also capture local dependencies.
\item \textbf{BERT4Rec} \cite{Sun2019bert} uses a masked item training scheme similar to masked language model order in NLP.
\item \textbf{VSAN} \cite{VSAN} introduces the self-attention network and variational autoencoder for sequential recommendation.
\end{itemize}

\textbf{\textit{3) Contrastive Learning Methods:} }
\begin{itemize}
\item \textbf{ACVAE} \cite{ACVAE} introduces adversarial training to sequence generative models and enables the model to generate high-quality augmented views for contrastive learning.
\item \textbf{DuoRec} \cite{DBLP:conf/wsdm/QiuHYW22} is the state-of-the-art SSL-based sequential method that employs a model-level augmentation approach based on dropout and a novel sampling strategy to construct contrastive self-supervised signals.
\item \textbf{ContrastVAE} \cite{ContrastVAE} is the first VAE-based work combining data augmentation with model augmentation for recommendation.
\end{itemize}

\subsection{Overall Performance (\textbf{RQ1})} 
Table \ref{Overall Comparison} shows the performance comparison of the proposed Meta-SGCL and other baseline methods on three real datasets. From the table, we can draw the following conclusions:

As can be seen from the table, traditional recommendation methods like Pop and BPR-MF exhibit the worst recommendation performance. However, GRU-based GRU4Rec consistently outperforms the non-sequential BPR-MF method, indicating that sequential modeling of user interaction sequences can effectively enhance recommendation performance. The Caser method, which utilizes convolutional neural networks to model sequence information, achieves similar recommendation performance to GRU4Rec.
The sequential recommendation method based on the attention mechanism has achieved strong recommendation performance. The SASRec method is the first work to use the attention mechanism in the sequence recommendation task. Compared with the traditional deep sequence-based model, its performance has been greatly improved, and the BERT4Rec method will learn more semantics, so as to achieve more good recommendation performance. Although the masked item prediction task can learn semantic features, it also introduces noise into the model, which cannot be well aligned with the sequence recommendation task to a certain extent.
Uncertainty modeling can well capture the diversity of user preferences. The VSAN model, on the one hand, can use VAE to model the user's uncertainty preference problem, and on the other hand, can use SAN to capture the dynamic changes of user preferences, and then achieve better recommendation performance than the above methods.

For the recent contrastive learning-based methods ACVAE, DuoRec, and ContrastVAE models, significant improvements are achieved over vanilla and sequential recommendation methods. The Clothing and ML-1M datasets achieve the strongest performance on the baseline DuoRec, while the Toys dataset achieves the strongest performance on the baseline ContrastVAE, and the Clothing and ML-1M datasets can obtain high-quality comparisons through model-based enhancements view, and the Toys dataset can only obtain high-quality comparative views through uncertainty modeling. Experimental results show the effectiveness of contrastive learning in sequence recommendation tasks.

% \begin{figure}[t]
% \centering
% \subfloat[The performance under different $\alpha$.]
% {\includegraphics[width=.5\linewidth]{fig/alpha_H10.pdf}%
%  \label{alphaH}}
% \subfloat[The performance under different $\alpha$.]
% {\includegraphics[width=.5\linewidth]{fig/alpha_N10.pdf}%
% \label{alphaN}}
% \\
% \subfloat[The performance under different $\beta$.]
% {\includegraphics[width=.5\linewidth]{fig/beta_H10.pdf}%
% \label{betaH}}
% \subfloat[The performance under different $\beta$.]
% {\includegraphics[width=.5\linewidth]{fig/beta_N10.pdf}%
% \label{betaN}}
% \\
% \subfloat[The performance under different $d$.]
% {\includegraphics[width=.5\linewidth]{fig/d_H10.pdf}%
% \label{dH}}
% \subfloat[The performance under different $d$.]
% {\includegraphics[width=.5\linewidth]{fig/d_N10.pdf}%
% \label{dN}}
% \caption{Influence of hyper-parameters $\alpha$, $\beta$ and $d$.}
% \label{hyper}
% \end{figure}
\begin{figure}[t]
\centering
\includegraphics[width=0.5\textwidth]{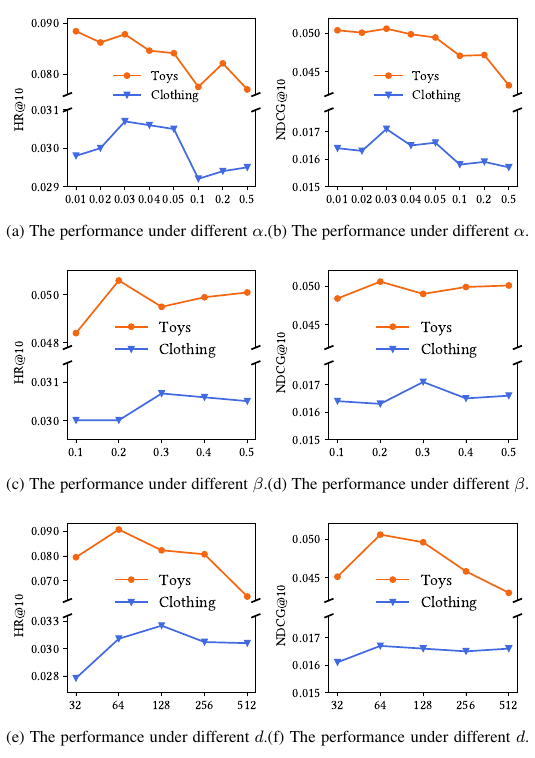}
%\vspace{-1cm}
\caption{Influence of hyper-parameters $\alpha$, $\beta$ and $d$.} 
\label{hyper}
\end{figure}

% Our model achieves the best recommendation performance. Benefiting from the operation of meta-optimized generating adaptive contrastive views, Meta-SGCL shows varied improvements across different datasets when evaluated on different metrics. For example, for the Clothing dataset, the improvement of Meta-SGCL over the strongest baseline DuoRec on the four evaluation indicators is 2.32$\%$-25.66$\%$. For the Toys dataset, the improvement of Meta-SGCL over the strongest baseline ContrastVAE on the four evaluation indicators is 14.74$\%$-19.34$\%$. For the ML-1L dataset, the improvement of Meta-SGCL on the four evaluation indicators compared with the strongest baseline DuoRec is 16.25$\%$-20.84$\%$. The main reasons are as follows: First, uncertainty modeling can well model the diversity of user preferences. Secondly, contrastive learning can well solve the problem of user data interaction data in sequence recommendation. Finally, meta-optimized two-step training strategy generates adaptive contrastive views for contrastive learning.
Our model, Meta-SGCL, achieves the best recommendation performance, and this can be attributed to the effective operation of meta-optimized generating adaptive contrastive views. Notably, Meta-SGCL demonstrates varied improvements across different datasets when evaluated on different metrics. For instance, on the Toys dataset, Meta-SGCL surpasses the strongest baseline ContrastVAE by 14.74$\%$ to 19.34$\%$ across the four evaluation indicators. Similarly, on the ML-1L dataset, Meta-SGCL achieves an improvement of 16.25$\%$ to 20.84$\%$ over the strongest baseline DuoRec across the four evaluation indicators. Several key reasons contribute to the superior performance of Meta-SGCL. First, uncertainty modeling can well model the diversity of user preferences. Secondly, contrastive learning can well solve the problem of user data interaction data in sequence recommendation. Finally, the meta-optimized two-step training strategy generates adaptive contrastive views for contrastive learning. In summary, the combination of uncertainty modeling, contrastive learning, and meta-optimized training contributes to the excellent performance of Meta-SGCL, making it a powerful and versatile model for sequence recommendation tasks across diverse datasets. 

\subsection{Effectiveness of Meta Optimization (\textbf{RQ2})}
We conduct multiple experiments based on Meta-SGCL to demonstrate the effectiveness of meta-optimized training. We first compare the performance of the joint learning training strategy with Meta-SGCL, and the results are shown in Figure \ref{2S}, which proved the meta-optimized two-step training strategy also plays a positive role in improving the performance of the model. From the figure, it can be seen that our Meta-SGCL model has achieved better recommendation performance than the joint learning training strategy on all datasets. The results prove that the meta-optimized two-step training strategy has played an important role in improving the performance of our model. The main reasons are as follows: the joint learning training strategy leads to the generation of contrastive views by the model in a random manner, which cannot ensure the quality of the generated contrastive views. The meta-optimized two-step training strategy can adaptively generate high-quality contrastive views according to downstream tasks.

 \begin{table}[t]
\renewcommand\arraystretch{1.2}
\centering
\caption{The performance achieved by Meta-SGCL variants on three datasets.}
\resizebox{0.5\textwidth}{!}{
\begin{tabular}{c|c|ccccc}
\hline
Dataset 
& Metric  & -$clkl$ & -$cl$  & -$kl$  & Meta-SGCL  \\ \hline
\multirow{4}{*}{Clothing} 
&HR@5     & 0.0168 & 0.0191 & 0.0190  & \textbf{0.0216} \\
&HR@10    & 0.0272 & 0.0264 & 0.0265  & \textbf{0.0309}  \\
&NDCG@5   & 0.0091 & 0.0132 & 0.0132  & \textbf{0.0142} \\
&NDCG@10  & 0.0124 & 0.0155 & 0.0156  & \textbf{0.0167} \\    \hline
\multirow{4}{*}{Toys} 
&HR@5       & 0.0429 & 0.0608 & 0.0587  & \textbf{0.0642} \\
&HR@10      & 0.0652 & 0.0858 & 0.0849  & \textbf{0.0907} \\
&NDCG@5     & 0.0248 & 0.0401 & 0.0392  & \textbf{0.0420} \\
&NDCG@10    & 0.0320 & 0.0482 & 0.0477  & \textbf{0.0506} \\      \hline
\multirow{4}{*}{ML-1M} 
&HR@5       & 0.1087 & 0.1748 & 0.1841  & \textbf{0.2387} \\
&HR@10      & 0.1904 & 0.2685 & 0.2748  & \textbf{0.3560} \\
&NDCG@5     & 0.0638 & 0.1153 & 0.1235  & \textbf{0.1622} \\
&NDCG@10    & 0.0910 & 0.1455 & 0.1528  & \textbf{0.1953} \\
\hline
\end{tabular}
}
\label{Ab}
\end{table}

\subsection{Ablation Study of Meta-SGCL (\textbf{RQ3})}
In this experiment, the effects of KL-divergence loss, contrastive learning loss, and two-step update strategy are evaluated. The variants are: 
(1) -$clkl$: Removing KL-divergence module and contrastive learning module, and using the SASRec model for sequential recommendation.
(2) -$cl$: Removing contrastive learning module, and using KL-divergence module for sequential recommendation.
(3) -$kl$: Removing the KL-divergence module and only using contrastive learning module for sequential recommendation.

The results are shown in table \ref{Ab} and the following conclusions can be drawn. Firstly, when both KL-divergence module and contrastive learning module are removed, the recommendation performance is the worst. The reason is that when these two modules are removed, our model degenerates into a simple SASRec, which only models the sequential correlation of sequences. 
On the contrary, contrastive learning can effectively mine the supervised signals of the data itself and uncertainty modeling can effectively capture the diversity of user preferences, so that our model can capture the dynamics and diversity of user preferences at the same time, and the results prove that the two modules effectiveness. Second, when removing the KL-divergence module or the contrastive learning module, the recommendation performance degrades to varying degrees. As can be seen from the table, both modules play equal roles in our model. For Meta-SGCL, no modules are removed, and it yields the best performance compared to all other methods.

\begin{table}[t]
\renewcommand\arraystretch{1.2}
\centering
\caption{Influence of the number of self-attention heads.}
\resizebox{0.5\textwidth}{!}{
\begin{tabular}{c|c|cccccc}    
\hline
Dataset  & $h$  & HR@5 &HR@10   &NDCG@5  &NDCG@10  \\ \hline
\multirow{5}{*}{Clothing} 
& 1    & 0.0212             & 0.0302 
       & \textbf{0.0145}    & \textbf{0.0174} \\
& 2    & \textbf{0.0216}    & \textbf{0.0309}      
       & 0.0142             & 0.0167   \\
& 4    & 0.0203 & 0.0294 & 0.0139 & 0.1690   \\
& 8    & 0.0200 & 0.0273 & 0.0137 & 0.0160   \\     \hline
\multirow{5}{*}{Toys} 
& 1    & 0.0586  & 0.0812 & 0.0392 & 0.0465   \\
& 2    & \textbf{0.0642}            & \textbf{0.0907}      
       & \textbf{0.0420}            & \textbf{0.0506}   \\
& 4    & 0.0551  & 0.0782 & 0.0388 & 0.0462   \\
& 8    & 0.0562  & 0.0779 & 0.0392 & 0.0462    \\   \hline
\end{tabular}}
\label{heads}
\end{table}

\subsection{Hyper-parameter Sensitivity (\textbf{RQ4})}
\subsubsection{Influence of the $\alpha$ to control the CL term}
We first investigate how the contrastive learning term of our proposed Meta-SGCL interacts with the sequential prediction term and KL term. 
Specifically, we explore how different $\alpha$ impact the recommendation performance. We keep other parameters fixed to make a fair comparison. Figures 5(a) and 5(b) show the evaluation results. Note that a larger value $\alpha$ contributes more heavily in the total. We observe that performance substantially deteriorates when $\alpha$ increases over a certain threshold. This observation implies that when contrastive learning loss dominates the learning process, it may decrease the performance on the sequence prediction task. We will analyze this impact carefully in future work. We can see that our framework requires a proper choice of the hyper-parameter $\alpha$ to reach its best performance, and a proper choice of $\alpha$ is around 0.03. 
 
\subsubsection{Influence of the $\beta$ to control the KL term}
We set $\beta$ to a fixed value from $\left \{ 0.1,0.2,0.3,0.4,0.5 \right \}$ for related experiments. Figures 5(c) and 5(d) display the experimental results under different $\beta$ on both datasets. We can observe that increasing the value of $\beta$ will improve the performance, and when $\beta$ exceeds a certain value, it will hurt the model performance. Appropriate selection of the value of $\beta$ can make our KL-divergence and recommendation tasks cooperate and enhance each other. In short, we recommend carefully tuning $\beta$ in the range $0.1\sim 0.5$. And we set the $\beta$ as 0.2 for our model on the Toys while setting 0.3 on the Clothing.

\subsubsection{Influence of the embedding dimension $d$} 
The embedding dimension has a great influence on the model performance. Therefore, we investigate the influence of the embedding dimension $d$ ranging from ${32, 64, 128, 256, 512}$ on Clothing and Toys datasets. We investigate the influence of the embedding size $d$ and report the results in Figures 5(e) and 5(f). Obviously, the high dimension can improve the performance of the network. This phenomenon is similar to the traditional latent factor model. But when the dimension exceeds a certain value, the results will no longer increase, and even begin to decline. This shows that it may cause overfitting when the embedding dimension of the latent factor is too large. Finally, when $d$ is 64, our model achieves the best recommendation performance. The embedding size is highly related to the model recommendation performance, and an appropriate embedding size can make our model achieve the best results.

\begin{table}[t]
\renewcommand\arraystretch{1.2}
\centering
\caption{Influence of the temperature parameter $\tau$.}
\resizebox{0.5\textwidth}{!}{
\begin{tabular}{c|c|cccccc}    
\hline
Dataset  & $\tau$  & HR@5 &HR@10   &NDCG@5  &NDCG@10  \\ \hline
\multirow{6}{*}{Clothing} 
& 0.05 & 0.0210          & 0.0302          & 0.0148          & \textbf{0.0177} \\
& 0.1  & \textbf{0.0218} & \textbf{0.0312} & \textbf{0.0144} & 0.0174          \\
& 0.5  & 0.0210          & 0.2970          & 0.0144          & 0.0172          \\
& 1    & 0.0216          & 0.0309          & 0.0142          & 0.0167          \\
& 2    & 0.0208          & 0.0291          & 0.0144          & \textbf{0.0177} \\
& 5    & 0.0201          & 0.0281          & 0.0140          & 0.0165          \\ \hline
\multirow{6}{*}{Toys} 
& 0.05 & 0.0562          & 0.0791          & 0.0396          & 0.0470          \\
& 0.1  & 0.0573          & 0.0803          & 0.0406          & 0.0480          \\
& 0.5  & 0.0569          & 0.0794          & 0.0402          & 0.0474          \\
& 1    & \textbf{0.0642} & \textbf{0.0907} & \textbf{0.0420} & \textbf{0.0506} \\
& 2    & 0.0565          & 0.0789          & 0.0393          & 0.0464          \\
& 5    & 0.0552          & 0.0744          & 0.0391          & 0.0453           \\ \hline
\end{tabular}}
\label{tua}
\end{table}

\subsubsection{Influence of the number of self-attention heads} 
Multi-head attention is to project Q, K, and V through multiple different linear transformations, and finally stitch together different attention results. We discuss the effects of multi-head in modeling sequence patterns. Figure \ref{heads} shows the experimental results, where $h$ represents the number of self-attention heads used for learning transition patterns. We can observe that when the number of heads is 2, the model performs best on the Clothing and Toys dataset. On the Clothing dataset, when $h$=1, the value in terms of NDCG@5 and NDCG@10 are the largest. It indicates that Meta-SGCL may not require too complex structures to model these relationships.

\subsubsection{Influence of the temperature parameter $\tau$}
The role of the temperature coefficient is to adjust the degree of attention to difficult samples: the smaller the temperature coefficient, the more attention is paid to separating this sample from the most similar other samples. $\tau$ plays a critical role in hard negative mining. Table \ref{tua} shows the curves of model performance w.r.t. different $\tau$. We can observe that: (1)  Increasing the value of $\tau$ (e.g., 5) will lead to poorer performance, which falls short in the ability to distinguish hard negatives from easy negatives. (2) In contrast, fixing $\tau$ to a too-small value (e.g., 0.05) will hurt the model performance, since the gradients of a few negatives dominate the optimization, losing the supremacy of adding multiple negative samples in the SSL objective. Hence, we suggest tuning $\tau$ in the range of $0.1 \sim 1.0$ carefully.

\subsubsection{Influence of the dropout rate}
Dropout means that during the training process, the neural network unit is temporarily dropped from the network according to a certain probability. It has been proven to be an effective dropout way benefits to prevent the overfitting of various neural networks. Therefore, we investigate the influence of the dropout rate ranging from 0 to 0.4 on two datasets. Table \ref{dr} displays the experimental results under different dropout rate settings on both datasets. It can be seen from the figure that when the dropout rate is set to 0 (that is, no neurons are dropped), the result is relatively poor. This proves that dropout is indeed effective in preventing overfitting. Moreover, for the Clothing and Toys datasets, when the dropout rate is set to 0.2, the best result can be obtained. Therefore, in our work, we set it to 0.2.  Finally, we can observe that the table shows the same trend: as the dropout rate increases, the performance of the model first improves and then declines or even declines sharply. This demonstrates that a proper dropout rate helps improve the expressive ability and generalization ability of the model. But when the dropout rate is too large, too many neurons are lost, which will limit the expression of the model.

\begin{table}[t]
\renewcommand\arraystretch{1.2}
\centering
\caption{Influence of the dropout rate.}
\resizebox{0.5\textwidth}{!}{
\begin{tabular}{c|c|cccccc}    
\hline
Dataset  & dropout rate  & HR@5 &HR@10   &NDCG@5  &NDCG@10  \\ \hline
\multirow{5}{*}{Clothing} 
& 0   & 0.0204          & 0.0295          & 0.0142          & 0.0171          \\
& 0.1 & 0.0206          & 0.0284          & \textbf{0.0146} & 0.0171          \\
& 0.2 & \textbf{0.0216} & \textbf{0.0309} & 0.0142          & 0.0167          \\
& 0.3 & 0.0209          & 0.0299          & 0.0144          & \textbf{0.0173} \\
& 0.4 & 0.0211          & 0.0289          & \textbf{0.0146} & 0.0172          \\ \hline
\multirow{5}{*}{Toys} 
& 0   & 0.0558          & 0.0781          & 0.0376          & 0.0448          \\
& 0.1 & 0.0569          & 0.0787          & 0.0395          & 0.0456          \\
& 0.2 & \textbf{0.0642} & \textbf{0.0907} & \textbf{0.0420} & \textbf{0.0506} \\
& 0.3 & 0.0576          & 0.0794          & 0.0397          & 0.0467          \\
& 0.4 & 0.0570          & 0.0763          & 0.0411          & 0.0473           \\ \hline
\end{tabular}}
\label{dr}
\end{table}

\subsubsection{Impact of the similarity function}
Cosine similarity could effectively measure the similarity of two vectors, regardless of the actual length of the two vectors. Even if the lengths of $A$ and $B$ are super short and super long, the cosine similarity of the two may also be 1. Cosine similarity only depends on the angle of two vectors, while dot product depends on both the angle and length of the two vectors. In the subsection, we experiment with the above two similarity functions. The specific results are shown in Table \ref{sim}. The Sports and Toys datasets have the best results on the dot product function on the two evaluation indicators, which shows that only the angle between the two vectors can describe the similarity of the two vectors well.
\begin{table}[t]
\renewcommand\arraystretch{1.2}
\centering
\caption{The performance achieved by Meta-SGCL variants on three datasets.}
\resizebox{0.5\textwidth}{!}{
\begin{tabular}{c|c|cccccc}    
\hline
Dataset  & similarity function  & HR@5 &HR@10   &NDCG@5  &NDCG@10  \\ \hline
\multirow{2}{*}{Clothing} 
& dot    & \textbf{0.0216}       & \textbf{0.0309}      & \textbf{0.0142}       & \textbf{0.0167} \\
& cos    & 0.0210       & 0.0294      & 0.0139       & 0.0166 \\     \hline
\multirow{2}{*}{Toys} 
& dot    & \textbf{0.0642}       & \textbf{0.0907}      & \textbf{0.0420}       & \textbf{0.0506}   \\
& cos    & 0.0563       & 0.0795      & 0.0379       & 0.0454    \\   \hline
\multirow{2}{*}{ML-1M} 
& dot    & \textbf{0.2387}       & \textbf{0.3560}      & \textbf{0.1622}       & \textbf{0.1950}   \\
& cos    & 0.1758       & 0.2750      & 0.1150       & 0.1420   \\   \hline
\end{tabular}}
\label{sim}
\end{table}

\subsection{Robustness to Noisy Data (\textbf{RQ5})}
\begin{figure}[t]
\centering
\includegraphics[width=0.68\textwidth]{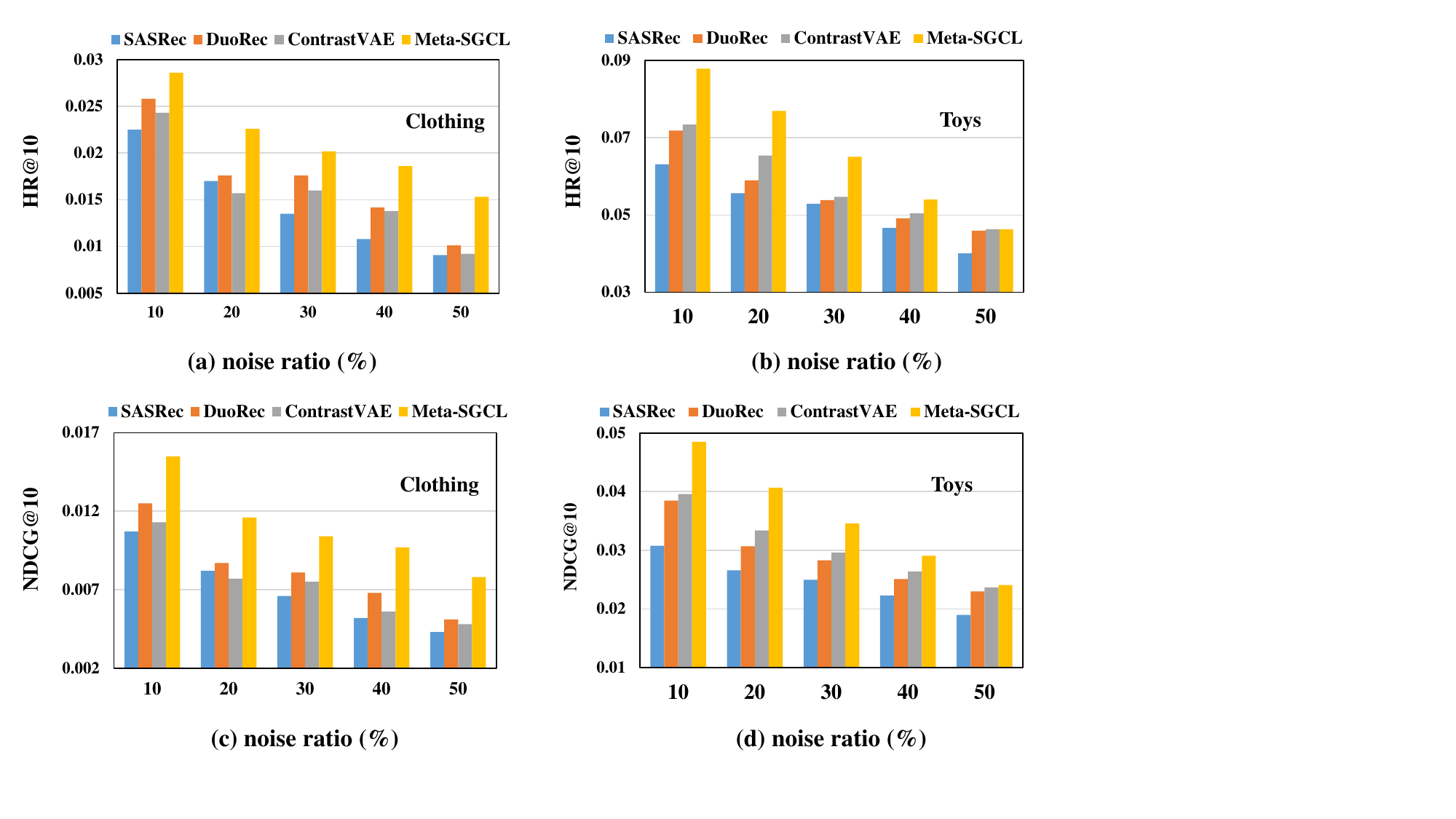}
\vspace{-1cm}
\caption{Model performance comparisons with respect to different settings of noise ratio on Toys and Clothing datasets.} 
\label{noise}
\end{figure}

To verify the robustness of Meta-SGCL against noisy interactions, we randomly add a certain proportion (10$\%$, 20$\%$, 30$\%$, 40$\%$ and 50$\%$) of negative items into the input sequences during training, and examine the final performance of Meta-SGCL and other baselines. From Figure \ref{noise}, we can see that adding noisy data deteriorates the performance of all models. By comparing SASRec and DuoRec, it can be seen that adding a self-supervised auxiliary task can significantly improve the model's robustness. By comparing Meta-SGCL and other models, it can be seen that our model also performs better than other models. Especially, with 10 noise proportion, our model can even outperform other models without noisy dataset. This is probably because the meta-optimization strategy can automatically generate high-quality augmented views, thus endowing the decoder with more robustness against noisy data.
% \begin{figure}[t]
% \centering
% \includegraphics[width=0.7\textwidth]{fig/noise.pdf}
% \vspace{-1cm}
% \caption{Model performance comparisons with respect to different settings of noise ratio on Toys and Clothing datasets.} 
% \label{noise}
% \end{figure}

\subsection{Visualization of Item Embedding (\textbf{RQ6})} 
\begin{figure}[t]
\centering
\vspace{-1cm}
\includegraphics[width=0.51\textwidth]{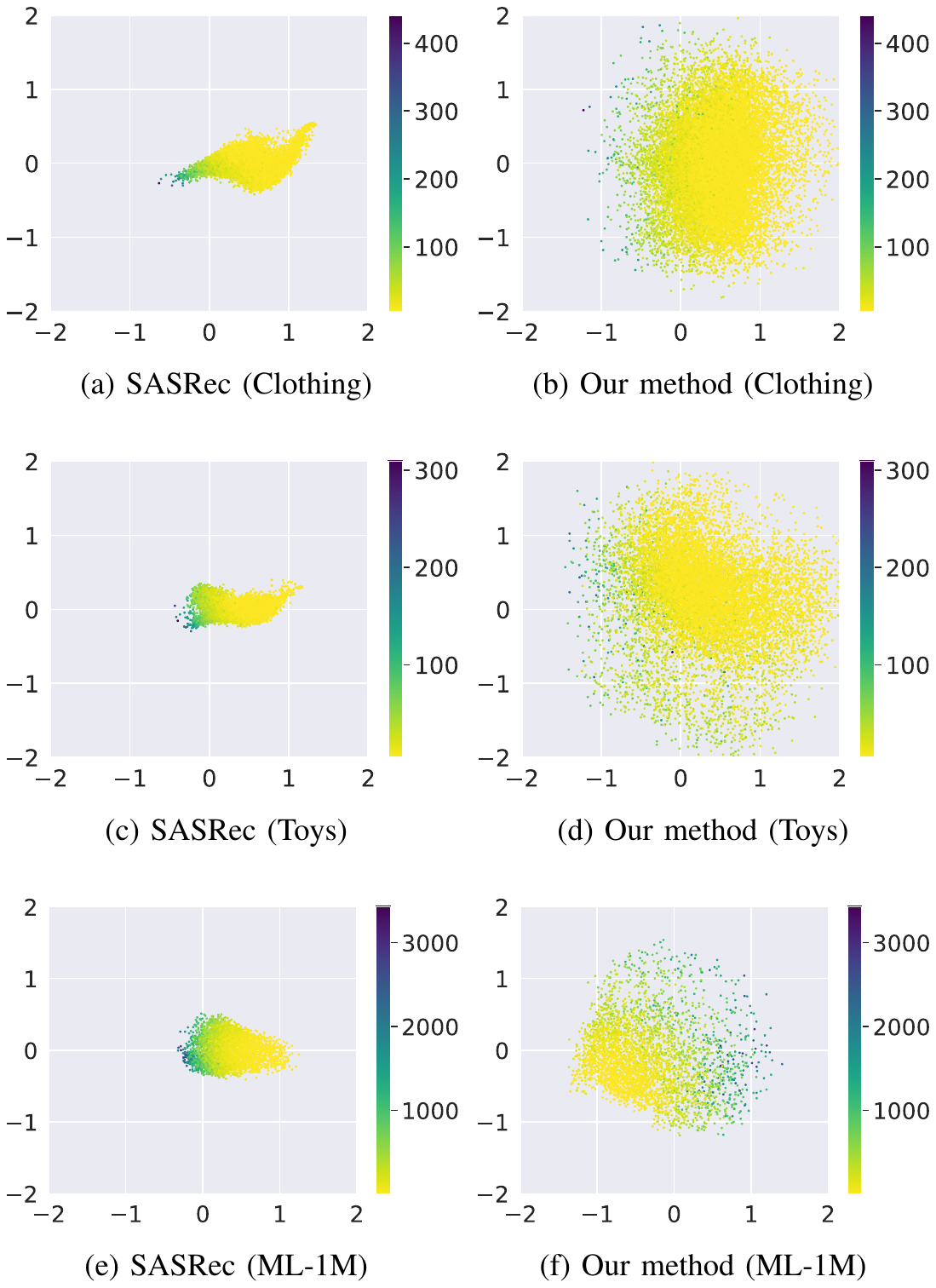}
\vspace{-1cm}
\caption{The item embedding of the three datasets on the two models (different colors indicate the different frequencies of items in the dataset).} 
\label{item embedding}
\end{figure}

To evaluate how contrastive learning affects the method, visualizations of the learned embedding matrix will be presented to help understand how contrastive learning improves performance. The result is shown in Figures \ref{item embedding}. For the SASRec method, the item embeddings for three different datasets are shown in Figures 7(a), 7(c) and 7(e). From these figures, it can be seen that SASRec will produce a narrow cone in the latent space and the distribution is relatively concentrated. For our Meta-SGCL method, the item embeddings for three different datasets are shown in Figures 7(b), 7(d) and 7(f). As can be seen from the figure, Meta-SGCL can produce higher quality embedding distributions mainly because it uses the Seq2Seq generator and contrastive learning. It can be concluded that, on the one hand, the Seq2Seq generator does not destroy the order of the sequence and can produce higher-quality augmentation views. On the other hand, the meta-optimized two-step training strategy adapts contrastive views for the different datasets, and thus the resulting sneak distribution is more uniform.

% \begin{figure}[t]
% \centering
% \subfloat[SASRec (Clothing) \label{Clothing1}]
% {\includegraphics[width=.48\linewidth]{fig/SAS_Clothing.pdf}}
% \subfloat[Our method (Clothing)\label{Clothing2}]
% {\includegraphics[width=.48\linewidth]{fig/our_Clothing.pdf}}
% \\
% \subfloat[SASRec (Toys) \label{Toys1}]
% {\includegraphics[width=.48\linewidth]{fig/SAS_Toys.pdf}}
% \subfloat[Our method (Toys)\label{Toys2}]
% {\includegraphics[width=.48\linewidth]{fig/our_Toys.pdf}}
% \\
% \subfloat[SASRec (ML-1M) \label{ML-1M1}]
% {\includegraphics[width=.48\linewidth]{fig/SAS_ML-1M.pdf}}
% \subfloat[Our method (ML-1M)\label{ML-1M2}]
% {\includegraphics[width=.48\linewidth]{fig/our_ML-1M.pdf}}
% \caption{The item embedding of the three datasets on the two models (different colors indicate the different frequencies of items in the dataset).}
% \label{item embedding}
% \end{figure}

\section{Related Work}
% In this section, we review the relevant methods in sequential recommendation, contrastive learning and meta-learning.

\subsection{Sequential Recommendation}
Sequential recommendation aims to predict the next item of the user's interest based on historical interaction sequences. Traditional methods for sequential recommendation adopt Markov Chains (MCs) to model the transition patterns between the items, such as FPMC \cite{DBLP:conf/www/RendleFS10} and Fossil \cite{Fossil}. 
Later, Recurrent Neural Network (RNN) \cite{DBLP:conf/emnlp/ChoMGBBSB14} and its variants \cite{srnn2016} are introduced into sequential recommendation to extract temporal transitions. Convolutional Neural Network (CNN) is also effective in modeling short-term dependencies for sequential recommendation, such as Caser \cite{tang2018caser}. 
More recently, Self-attention Network becomes the mainstream architecture for sequential recommendation due to its flexibility of capturing both global and local features, serving as the backbone of many sequential models, such as SASRec \cite{kang18attentive}, BERT4Rec \cite{Sun2019bert} and VSAN \cite{VSAN}. However, these works are often limited by data sparsity and uncertainty in real-world applications.

\subsection{Contrastive Learning}
Contrastive Learning (CL) is a generalized paradigm that requires the model to predict parts of the unobserved data from other observed parts. 
It can be classified into two categories, i.e., the generative CL and the contrastive CL. 
 The successful applications of SSL in computer vision \cite{chen2020simple} 
and natural language processing \cite{gao2021simcse} 
motivate researchers to apply CL in recommender systems. For example, BERT4Rec \cite{Sun2019bert} applies  the generative SSL in sequential recommendation by formulating sequential recommendation as a masked item prediction task. Other examples of the generative SSL include \cite{ACVAE} and ContrastVAE \cite{ContrastVAE}, which trains a VAE) to map the user sequence into latent variables and then reconstruct the user sequence through variational inference. As for the contrastive SSL, many works try to apply augmentations to the original sequence and then maximize the agreements between a pair of augmented sequences, such as CL4SRec \cite{DBLP:journals/corr/abs-2010-14395}, CoSeRec \cite{DBLP:journals/corr/abs-2108-06479} and DuoRec \cite{DBLP:conf/wsdm/QiuHYW22}. S$^{3}$-Rec \cite{DBLP:conf/cikm/ZhouWZZWZWW20} devises self-supervised objectives to fuse contextual information into sequential recommendation. Different from these works that focuses on the generative SSL or the contrastive SSL, this paper achieves a synergy between both the  generative and the contrastive SSL to alleviate the issues of data sparsity.
\subsection{Meta-Learning}
Meta-Learning trains a meta-learner that automatically learns the optimal algorithm for different tasks \cite{MAML}. such as CML \cite{dong2022cml}  in NLP, which combines contrastive and meta-learning techniques to enhance text feature representations and generate confidence scores for label quality.
In recommender systems, meta-learning is often adopted to solve the cold-start issue, where only a few interaction records are available for cold-start users. For example, MAMO \cite{MAMO} designs two memory matrices and a meta-optimization approach to encourage personalized parameter initialization for each user. MFNP \cite{MFNP} optimizes both user-specific and region-specific modules with meta-learning algorithms for cold-start POI recommendation. Mecos \cite{Mecos} deals with the cold-start items in sequential recommendation via meta-learning based gradient descent approach. Different from these works, this paper trains the sequence encoder as a meta-learner that learns to update the parameters in a personalized style.

\section{CONCLUSION}
In this paper, we propose Meta-optimized Joint Seq2Seq Generator and Contrastive Learning for Sequential Recommendation (Meta-SGCL), which applies the meta-optimized two-step training strategy to adaptive generate contrastive views. This is the first work to use the Seq2Seq generator to generate contrastive views in sequence recommendation and optimizes the model with the novel meta-optimized two-step update strategy. Extensive experimental results show that the proposed method outperforms the state-of-the-art sequential recommendation models. In addition, due to the generalization of our framework, Meta-SGCL can be applied to many other recommendation models and further improve their performance.
The future research directions include exploring different view generators, enabling online learning and real-time recommendations, enhancing model interpretability, considering privacy and fairness considerations.

%\section*{Acknowledgement}
%This research was partially supported by the NSFC (62376180, 62176175), the major project of natural science research in Universities of Jiangsu Province (21KJA520004), Suzhou Science and Technology Development Program(SYC2022139), the Priority Academic Program Development of Jiangsu Higher Education Institutions and the Exploratory Self-selected Project of the State Key Laboratory of Software Development Environment.
% Create an appendix in LaTeX
\bibliographystyle{IEEEtran}
\bibliography{main}

\end{document}